\documentclass[reprint,superscriptaddress,amsmath,amssymb,aps]{revtex4-2}
\usepackage{paper_style}
\myexternaldocument{supplement}
\begin{document}
\title{Pipeline quantum processor architecture for silicon spin qubits}
\newcommand\QMT{\,Quantum Motion, 9 Sterling Way, London N7 9HJ, United Kingdom}
\newcommand\UCL{\,London Centre for Nanotechnology, University College London, London WC1H 0AH, United Kingdom}
\newcommand\Oxford{\,Department of Materials, University of Oxford, Parks Road, Oxford OX1 3PH, United Kingdom}
\author{S. M. Patom{\"a}ki}
\affiliation{\QMT}
\affiliation{\UCL}
\email{ucapsmp@ucl.ac.uk}
\author{M. F. Gonzalez-Zalba}
\affiliation{\QMT}
\author{M. A. Fogarty}
\affiliation{\QMT}
\author{Z. Cai}
\affiliation{\Oxford}
\affiliation{\QMT}
\author{S. C. Benjamin}
\affiliation{\QMT}
\affiliation{\Oxford}
\author{J. J. L. Morton}
\affiliation{\QMT}
\affiliation{\UCL}
\date{\today}
\begin{abstract}
Noisy intermediate-scale quantum (NISQ) devices 
seek to achieve quantum advantage over classical systems without the use of full quantum error correction.
We propose a NISQ processor architecture using a qubit `pipeline' in which all run-time control is applied globally, reducing the required number and complexity of control and interconnect resources. 
This is achieved by progressing qubit states through a layered physical array of structures which realise single and two-qubit gates.
Such an approach lends itself to NISQ applications such as variational quantum eigensolvers which require numerous repetitions of the same calculation, or small variations thereof.
In exchange for simplifying run-time control, a larger number of physical structures is required for shuttling the qubits as the circuit depth now corresponds to an array of physical structures. However, qubit states can be `pipelined' densely through the arrays for repeated runs to make more efficient use of physical resources. 
We describe how the  
qubit pipeline can be implemented in a silicon spin-qubit platform, to which it is well suited to due to the high qubit density and scalability. 
In this implementation, we describe the physical realisation of single and two qubit gates which represent a universal gate set that can achieve fidelities of $\mathcal{F} \geq 0.9999$, even under typical qubit frequency variations. 
\end{abstract}
\maketitle
\section{Introduction} 
\label{section:introduction}
Fault-tolerant quantum computers offer profound computational speed-ups across diverse applications, but are challenging to build. However, even in the near term without full error correction, quantum computers have the potential to offer improvements over classical computing approaches in run time scaling~\cite{preskill2012quantum} and energy consumption for certain tasks~\cite{lloyd2000ultimate}. 
There is a rich and diverse array of schemes for realising quantum computation, each formally equivalent in computational power~\cite{mizel2007simple,nielsen2006cluster,deutsch1989quantum}, with important differences with regards to practical realisations.
In the gate-based approach, a quantum algorithm is expressed as a quantum circuit consisting of a series of quantum logic gates. Typically, such gates are applied to a stationary array of qubits (for example through electromagnetic waves or optical pulses), relying on the delivery of a complex series of accurate, quasi-simultaneous control pulses to each qubit. This can lead to practical challenges ranging from cross-talk in the control pulses between nearby qubits~\cite{undseth2023nonlinear,gonzalez2021scaling} to increased demands on digital-to-analogue converters (DACs) particularly when fully integrating control systems with a cryogenic quantum chip. 
\par
To mitigate the practical challenges associated with high-density control electronics, global qubit control schemes have been explored~\cite{benjamin2000schemes,benjamin2001quantum,fitzsimons2006globally}.
However, these approaches require a precision in the position and homogeneity of qubit structures, as well as the control pulses, to a degree that is technically challenging with available technology. Alternatively, local addressing can be used to bring qubits into resonance with globally applied control fields to create an effective local control~\cite{laucht2015electrically,wolfowicz2014conditional}. 
However, this approach still requires fast run-time control for the addressing~\cite{hansen2021pulse}.
A second strategy is to accept the cost of a much lower effective qubit density in exchange for mitigating effects such as cross-talk --- in such a distributed model of quantum computing~\cite{lim2005repeat} qubits, or small qubit registers, are well-separated and interfaced by a common entangling (typically photonic) mode.  
While such hybrid matter/photon systems  have been successfully realised in several platforms~\cite{moehring2007entanglement,campagne2018deterministic,bernien2013heralded}, high gate speeds compatible with the demands of NISQ algorithms~\cite{cai2020resource} may be difficult to achieve.
\par
A further consideration when designing NISQ hardware architectures is that NISQ algorithms based on (e.g.) variational approaches~\cite{cai2020resource,cade2019strategies} require multiple repetitions of the same quantum circuit --- or simple variations thereof --- to be performed. In turn this requires identical, or similar, sequences of local control fields to be repetitively applied to the qubits, presenting an opportunity for more efficient hardware implementations.  
In this Article, we propose a quantum computing architecture for implementing quantum circuits in which all runtime control is applied globally, and local quantum operations such as 1-qubit (1Q) and 2-qubit (2Q) gates are `programmed' into the array in advance. This is achieved by shuttling qubit states through a grid of gated structures which have been electronically configured to realise specific gates.
\par
In such an approach, each layer of gates in the original quantum circuit corresponds to a one-dimensional array of structures, such that the scheme is more demanding in terms of physical resources on a chip for confining qubits. However, when applying multiple repetitions of the same circuit by \emph{pipelining} --- i.e.\ running distinct, staggered layers of qubits through the array simultaneously --- the physical resource efficiency becomes broadly equivalent to more conventional approaches, combined with potential practical benefits.
\par
Below, we introduce the concept of a qubit pipeline in more detail in Sec.~\ref{section:qubit_pipeline}, before focusing on a potential implementation for the silicon metal-oxide-semiconductor (SiMOS) electron spin qubit platform in Sec~\ref{section:implementation_with_nanogrid}. We then estimate the expected improvement in run time for an example algorithm in Sec.~\ref{section:application_as_bespoke_NISQ_hardware}.
In particular, in Sec.~\ref{section:shuttling}, we outline a scheme for synchronous shuttling, and in Sec.~\ref{sec:initialisation_readout_preconfiguration}, we discuss initialisation, readout, and parallelised pre-configuration, mostly outlining established methods.  
In Sections~\ref{section:Z_rotation_gate}-\ref{section:transversal_qubit_control}, we show how to realise an universal gate set for the pipeline in the silicon electron spin platform including: single-qubit $Z$-rotations using local, voltage-controllable $g$-factor Stark shifts, globally-applied $\sqrt{X}$ operations enabled by $B_{1}$-drive frequency binning and 2Q SWAP-rotation gates using the native interaction of nearest neighbour exchange. Each of the above gate implementations is designed to accommodate natural variations, such as random $g$-factor differences, across the QDs while enabling synchronized operation and thus fixed gate times. 
\section{Qubit pipeline} 
\label{section:qubit_pipeline}
\begin{figure*}
\includegraphics[]{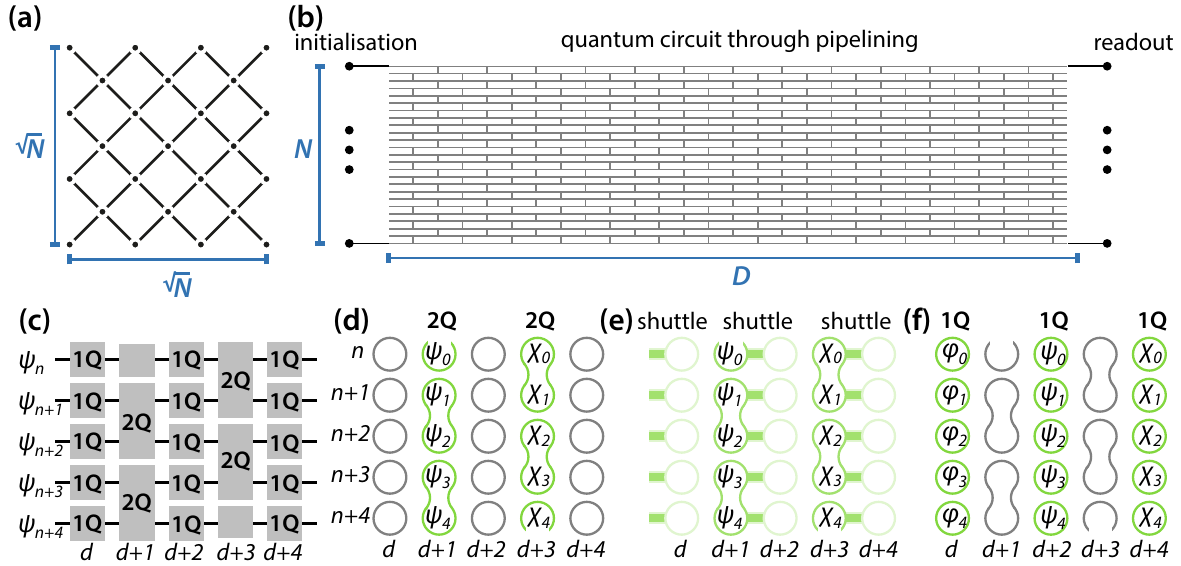}
\caption{
\label{fig:pipeline_concept}
\textbf{The qubit pipeline.} 
\textbf{(a)} In a typical $N$-qubit solid state quantum processor, qubits reside at fixed spatial locations (e.g. on a $\sqrt{N} \times \sqrt{N}$ grid) with nearest-neighbour connectivity. 
\textbf{(b)} The qubit pipeline is a weaved grid in which $N$ qubits are shuttled through $D$ locations where fixed single- (1Q) or two-qubit (2Q) logic gates are implemented.
Vertical lines indicate 2Q couplers, while horizontal lines indicate shuttling couplers. At runtime, qubits are initialized at one end, synchronously shuttled through the pipeline, and read out on the opposite end.
\textbf{(c)} An example quantum circuit diagram, where an algorithm is decomposed into alternating steps of 1Q and 2Q gates. 
\textbf{(d)}-\textbf{(f)} 
The qubit pipeline contains physical locations which have been configured to implement 1Q (circles) and 2Q (connected circles) gates. Different qubit arrays (first $(\chi_{0}, \chi_{1}, ...)$, then $(\psi_{0}, \psi_{1}, ...)$, $(\varphi_{0}, \varphi_{1}, ...)$) can be piped sequentially through the structures, each representing one execution of the configured quantum circuit.
}
\end{figure*}

For solid-state quantum processors the qubit array may consist of a two-dimensional lattice with nearest-neighbour couplings (see Fig.~\ref{fig:pipeline_concept}\textbf{(a)}).
The qubit pipeline approach replaces the two-dimensional lattice of stationary qubits with an $N \times D$ grid (see Fig.~\ref{fig:pipeline_concept}\textbf{(b)}), such that $N$ qubits are arranged in a one-dimensional array and propagate through the grid to perform a quantum circuit of depth $D$. 
Each column therefore represents a single `time step'  of quantum logic gates in the corresponding quantum circuit while each row of structures forms a `pipe' of computational length $D$ along which a qubit travels. Multiple qubits can simultaneously travel through each pipe, at different stages, in order to more efficiently use the physical resource. 
\par

\par
Run-time operation begins by initializing a one-dimensional array of $N$ qubit states on the input edge. This initialised array is synchronously pushed through $D$ structures that have been preconfigured to perform the single- and two-qubit gates in the desired quantum circuit. Qubit states are read out on the opposite output edge. Between initialization and readout, operations on the qubit array alternate between synchronous one- and two-qubit gate steps and shuttling steps. 
Figures~\ref{fig:pipeline_concept} \textbf{(c)}-\textbf{(f)} illustrate the equivalence between an example quantum circuit (Fig.~\ref{fig:pipeline_concept} \textbf{(c)}) and time-evolution on the qubit pipeline (Figs.~\ref{fig:pipeline_concept} \textbf{(d)}-\textbf{(f)}). In this way, local qubit control of the typical qubit grid is replaced by a combination of global control (at least, with widely shared control lines) to push the qubit states from start to finish, combined with quasi-statically tuneable elements within the larger number of physical structures used to define the quantum circuit prior to running it. 
\par
We assume that while the parameters of the quantum logic gates can be tuned in-situ (e.g.\ electrically), the type of gate performed in each cell of the array is defined when fabricating the quantum processor. Indeed, though more generally reconfigurable implementations may be possible (see e.g.\ Supplementary Sec.~\ref{sec:pixelgrid}) and offer potential efficiencies, a suitably chosen pattern of gates is sufficient for universal quantum computation, subject only to the constraints of the qubit number and circuit depth. 
Assuming the quantum processor is restricted to a two-dimensional topology, the pipeline approach presented here is limited to a linear qubit array. Many quantum algorithms can be mapped onto such a linear array without significant loss of efficiency, such as the variational quantum eigensolver for the Fermi-Hubbard model, a promising task that could be implemented on NISQ hardware~\cite{cai2020resource, cade2019strategies,kivlichan2018quantum,onedVsTwodEncoding}.
\par
Conventional approaches to operate an $N$ qubit processor with tuneable nearest neighbour couplings demand at least $\mathcal{O}(N)$ fast signal generators for single-qubit control, for qubit-qubit couplings, and for state readout. In contrast, the pipeline scheme presented here may require a constant (even just one) number of fast pulse generators, utilized for shuttling the qubit states through the grid in a manner analogous to a charge-coupled device. 
For example, three waveforms biasing columns $d\ \mathrm{mod}\ 3 = 0$, $d\ \mathrm{mod}\ 3 = 1$, and $d\ \mathrm{mod}\ 3 = 2$, create a local potential minimum for a qubit which can then be driven forward through the array~\cite{seidler2022conveyor}. 
Implementing such a shuttling scheme, synchronicity is achieved if gate steps take an equal amount of time for all qubits. For example, for logic gates expressed as $\exp(-i \omega \tau \sigma)$ for some single- or two-qubit operator $\sigma$, we fix a common $\tau$ but select $\omega$ to control the amount of rotation and thus distinguish the logic operations. The gate speed $\omega$ is varied using parameters such as dc gate voltages and magnetic field amplitudes at the preconfiguration stage. 
In principle, different columns of gates (e.g. all single-qubit gates, or all two-qubit gates) could have different durations, at the cost of a more complex shuttling pulse sequence. The three-column approach to shuttling represents the maximum density with which different sets of qubits can be pipelined through the circuit. 
\par
Due to the statistical nature of measurement in quantum mechanics, several types of quantum algorithms, such as those that yield an expectation value following some quantum circuit~\cite{mcardle2020quantum,harrow2009quantum}, must be run a large number of times to obtain a meaningful result.
Such algorithms are well-suited to the pipeline approach as multiple, independent qubit arrays can be pushed through the pipeline simultaneously, enabling multiple circuit runs in parallel. 
The maximum density of independent logical instances of a quantum circuit which can be pipelined through the structure is determined by the physical constraints of ensuring forward shuttling and avoiding unwanted interactions between qubits. Furthermore, if the duration of either the initialisation or readout stage is greater than that of the 1Q/2Q gates, this density must be further reduced. Exploiting such pipelining, the algorithm runtime is proportional to $(D + N_r) \tau$ instead of $D N_r \tau$, providing a significant speedup for circuits with large number of repetitions $N_r$, or large depths. Here, $\tau$ is the timescale of the longest operation: a qubit gate, readout, or initialization. Table \ref{tab:lattice_pipeline_comparison} summarises these main differences between a qubit lattice and the qubit pipeline. 
\par
\begin{center}
\begin{table}
\renewcommand{\arraystretch}{1.8}
\setlength{\tabcolsep}{1.5ex}
\begin{tabular}{lll}
\toprule
&
\textbf{Lattice}
&
\textbf{Pipeline}
\\ \midrule
\makecell[l]{\textbf{Processor} \vspace{-0.25em}
             \textbf{size}}
&
$\sqrt{N} \times \sqrt{N}$\ \ \ 
&
$N \times D$
\\ 
\makecell[l]{\textbf{Run-time} 
             \textbf{control} \vspace{-0.25em}
             \textbf{resources}} 
&
$N$
&
constant
\\ 
\makecell[l]{\textbf{Circuit} 
             \textbf{decomposition}}
& 
flexible
&
limited
\\ 
\makecell[l]{\textbf{Runtime} 
             \textbf{scaling}}
&
$D N_{\mathrm{r}}$
&
$D + N_{\mathrm{r}}$
\\ \bottomrule
\end{tabular}
\caption{
\label{tab:lattice_pipeline_comparison}
\textbf{Comparison between the quantum lattice and the pipeline processors.}
Here, $N$ is the number of qubits, $D$ is the maximum circuit depth, and $N_{\mathrm{r}}$ is the number of repetitions. 
}
\end{table}
\end{center}
\section{Implementation with silicon quantum dots} \label{section:implementation_with_nanogrid}
We now analyse a hardware implementation well-suited to the qubit pipeline paradigm, i.e. qubits based on single electron spins trapped in silicon-based QDs. 
Silicon spin qubits can achieve high density due to their small footprint of $\mathcal{O}(50\times 50\ \mathrm{nm}^{2})$, and can leverage the state-of-the-art nanoscale complementary metal-oxide-semiconductor (CMOS) manufacturing technology used in microprocessors~\cite{gonzalez2021scaling,li2018crossbar,vandersypen2017interfacing,boter2022spiderweb}. 
Quasi-CMOS-compatible electron spin qubits can be patterned as gated planar MOS devices~\cite{li2020flexible,veldhorst2014addressable,lawrie2020quantum}, confining electrons at the Si-SiO$_2$ interface below the gates, or using Si/SiGe heterostructures~\cite{lawrie2020quantum,xue2022quantum}. 
Typical values for single- and two-qubit gate fidelities measured so far in such systems include 99.96\% and 99.48\% in SiMOS~\cite{yang2019silicon, tanttu2023stability} and 99.9\% and 99.5\% in Si/SiGe~\cite{yoneda2018quantum,xue2022quantum, mills2022two}.
Coherent spin shuttling has been demonstrated at a transfer fidelity of 99.97\% for spin eigenstates, and $98$\% for spin-superposition states~\cite{yoneda2021coherent, noiri2022shuttling}, while SWAP gates have also be shown to transport arbitrary qubit states with fidelity up to 84\%~\cite{sigillito2019coherent}. 
In addition, silicon on insulator (SOI) nanowire and fin field-effect transistors (nwFET and finFET) devices~\cite{corna2018electrically,camenzind2021spin,zwerver2022qubits} have been used to confine spin qubits within etched silicon structures, and have been proposed for sparsely-connected two-dimensional qubit architectures~\cite{crawford2022compilation}. These devices typically show a large electrostatic gate control of the QDs (larger so-called lever arms $\alpha$) which is advantageous for reflectometry readout techniques, or spin-photon coupling~\cite{vigneau2022probing, burkard2020superconductor}. 
\par
\begin{figure}
\includegraphics[]{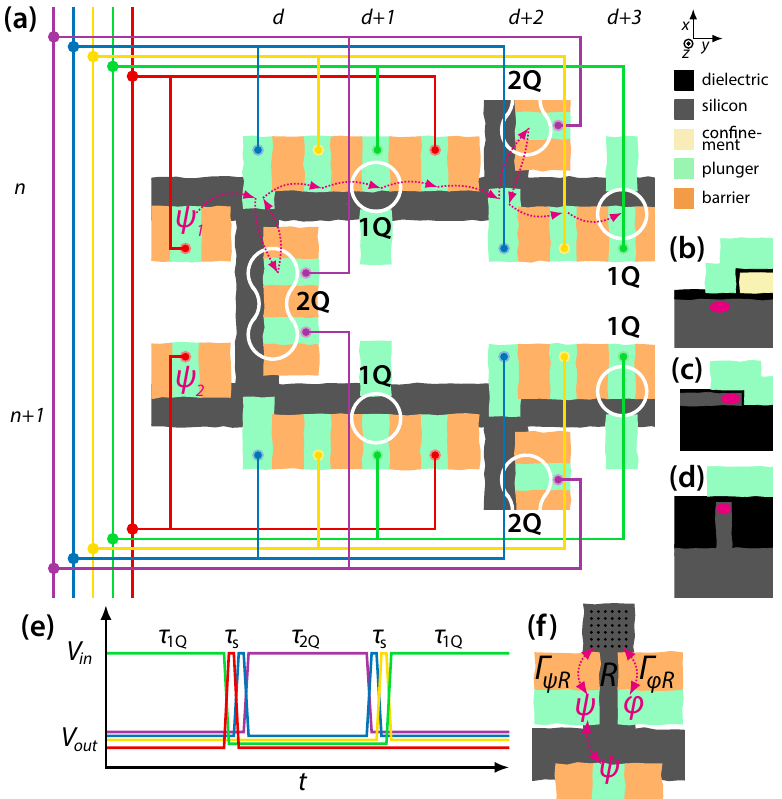}
\caption{
\label{fig:nanogrid_layout}
\textbf{Silicon quantum dot pipeline unit cell.} 
\textbf{(a)} Unit cell of a qubit pipeline, realised as a weaved grid of silicon channels (dark gray grid) which may be defined by etching or electrostatically by depletion gates (not shown). Overlapping metal gates (coloured rectangles) are used to confine, shuttle and manipulate electron spin qubits within the channels.
Connectivity for five-stage shuttling is shown as colored lines where all sites mod $5$ (for example, those controlling $\psi_{1}$ and $\psi_{6}$) are connected to the same voltage source.
\textbf{(b)}-\textbf{(d)} Side views of different gate stacks which could be used in this implementation, including  \textbf{(b)} planar MOS, \textbf{(c)} SOI nanowire, and \textbf{(d)} finFET. Quantum dots (magenta blobs) are confined using etched silicon or confinement gates (light yellow). 
\textbf{(e)} Sketch of the shuttling pulse sequence (relative pulse durations not to scale). Voltages $V_{\mathrm{in}}$ ($V_{\mathrm{out}}$) are those at which single (zero) electron occupancy of the QD becomes the ground state. Single- and two-qubit gates are separated by short shuttling steps -- in general, the number of shuttling steps depends on the exact gate layout, and depend on e.g. footprints required for routing. 
\textbf{(f)} Structures for local electron reservoirs (R), for spin readout (with an auxiliary state $\varphi$), and hence for preconfiguration of quantum states $\psi$. The operation of these structures, which can be placed along pipes between $d+1$ and $d+2$ (see panel \textbf{(a)}), is discussed further in the text. 
}
\end{figure}

To realise the qubit pipeline, we propose a sparse two-dimensional quantum dot(QD) array, which we refer to as a nanogrid (see Fig.~\ref{fig:nanogrid_layout} \textbf{(a)}). The nanogrid is a weaved grid of silicon `channels' in which QDs can be formed (e.g.\ etched silicon in the case of nwFET and finFET approaches, or through confining depletion gates in the case of planar MOS). Metal gates (green and orange shapes) for forming QDs are placed along and over the exposed Si to form locally one-dimensional QD arrays, which make 90-degree angles at T-junctions to join neighbouring pipes. Such weaving is required for entangling all the qubits from different pipes. Extra control from a second gate layer aids tuning all QDs to nominally identical setpoints despite variability in e.g. charging energies~\cite{yang2020quantum,bavdaz2022quantum}. The metal gates are routed with vias. The overall architecture can be realised in a variety of silicon QD platforms, including planar MOS, SOI, and finFETs, with cross-sections illustrated in Figures~\ref{fig:nanogrid_layout} \textbf{(b)}-\textbf{(d)}. In addition, a similar layout can be used in a Si/SiGe architecture or indeed other types of electron or hole semiconductor spin qubits.  
\par 

\par
\subsection{Shuttling}
\label{section:shuttling}
In the qubit pipeline, we shuttle electrons between different columns $d$ where logical 1Q and 2Q gate operations are performed.
Electrons can be shuttled from one QD to another by inverting the biasing between gates, i.e. by pulsing over the inter-dot charge transition with DQD charge occupancies $(1,0)\to(0,1)$~\cite{baart2016single,baart2016nanosecond, zwerver2022shuttling}. This scheme is also referred to as bucket-brigade shuttling. 
\par
The shuttling time $\tau_{s}$ is determined by the inter-site tunnel coupling frequency $t_{ij}/h$ (typically $1-20$ GHz in two-layer QD arrays using barrier gates~\cite{baart2016nanosecond}) and the electron temperature, which affects charge relaxation rates. 
Charge shuttling errors are expected to be minimized when the pulsing rate is slow compared to the inter-dot tunnel coupling~\cite{mills2019shuttling,yoneda2021coherent}, where non-adiabatic Landau-Zener (LZ) transitions are minimal. 
At tunnel coupling $t_{ij}/h = 20$ GHz, the ramp can be performed adiabatically ($P_{\mathrm{LZ}} < 10^{-4}$) with shuttling times of $9.1$ ns or more (see Supplementary Sec.~\ref{sec:minimizing_shuttling_errors}). 
Electron charges and spins have been demonstrated to shuttle reliably over these timescales~\cite{mills2019shuttling,yoneda2021coherent}, so we take $10$ ns as the range of target shuttling times $\tau_{s}$. The exact shuttling time should be determined to avoid LZ transitions to excited states, such as valley-orbit states. The systematic $Z$-phases arising from $g$-factor differences between sites can be accounted for as part of single-qubit control, as discussed in Sec.~\ref{section:Z_rotation_gate}. 
\par
The pulse sequence depicted in Fig.~\ref{fig:nanogrid_layout} 
\textbf{(e)} can realise the shuttling and logic gate dynamics modulation. 
To this end, each gate is routed to an ac + dc voltage combining circuit. The ac nodes from gates from a single column, and mod $5$ steps, are combined at interconnect level with a power splitter, whereas each gate receives an individual dc bias. Bias tees at gate nodes enable applying dc biases from individual dc sources. See Supplementary Sec.~\ref{sec:control_line_footprints} for footprint estimates.
As there are signals with three different periods, we could employ e.g. phase modulation or partially digital signal processing to generate the five shuttling biases, and hence shuttling biases for the entire pipeline with only three voltage pulse generators.
We can fill the pipeline up to every fifth physical gate column of QDs single electrons, which we refer to as maximal filling. 
\subsection{Initialization, readout and pre-configuration}
\label{sec:initialisation_readout_preconfiguration}
Initialisation and readout is analogous to a shift register for single-electron spins, where electrons are moved through a pipe and electron reservoirs are located at the input and output ends of the array.  
Preconfiguration of the unit cells can be done locally, utilizing initialization and readout nodes like the one depicted in Fig.~\ref{fig:nanogrid_layout} \textbf{(f)}. 
Logical ground states $\ket{ \hspace*{-0.25em} \downarrow_{q} }$ are initialised using spin-dependent reservoir-to-dot tunneling~\cite{elzerman2004single}. The fidelity is typically determined by the relative magnitude of the spin Zeeman energy splitting and $k_BT$ where $T$ is the temperature of the electron reservoirs and $k_B$ is Boltzmann's constant. For example, at $B_{0} = 1$ T, $\mathcal{F} \geq 0.9999$ at $T \lesssim 73$ mK.
The initialization time is determined by the reservoir-to-dot tunnelling time $1/\Gamma_{\psi \mathrm{R}}$, which can be controlled with a barrier gate, and can reach tens of GHz~\cite{oakes2023fast}. 
At higher electron reservoir temperatures, required fidelities could be obtained using real-time monitoring of the qubit through a negative-result measurement at the expense of longer initialization times~\cite{johnson2022beating}.  
\par
For readout, we employ the so-called Elzerman readout, where we detect spin states using the reverse of the spin-dependent tunneling described above, detected by a capacitively-coupled charge sensor, such as a single-electron box~\cite{borjans2021spin, ciriano2021spin}, at site $\varphi$ (see Fig.~\ref{fig:nanogrid_layout} \textbf{(f)}).  
This method is estimated to yield a spin readout fidelity $\mathcal{F} \geq 0.993$ in $\leq 4\ \upmu$s~\cite{oakes2023fast}, while advances in resonant readout techniques could help increase $\mathcal{F} \geq 0.9999$ in 50 ns~\cite{von2023multi}. 
We utilize Elzerman readout over Pauli spin blockade or parity readout, which are based on projections in the singlet-triplet basis, since we prefer to operate at relatively high $B_{0}$ fields, $B_{0} \approx 1$ T.
\subsection{\textit{Z}-rotation gates with Stark-shifted $g$-factors}
\label{section:Z_rotation_gate}

\begin{figure*}
\includegraphics[]{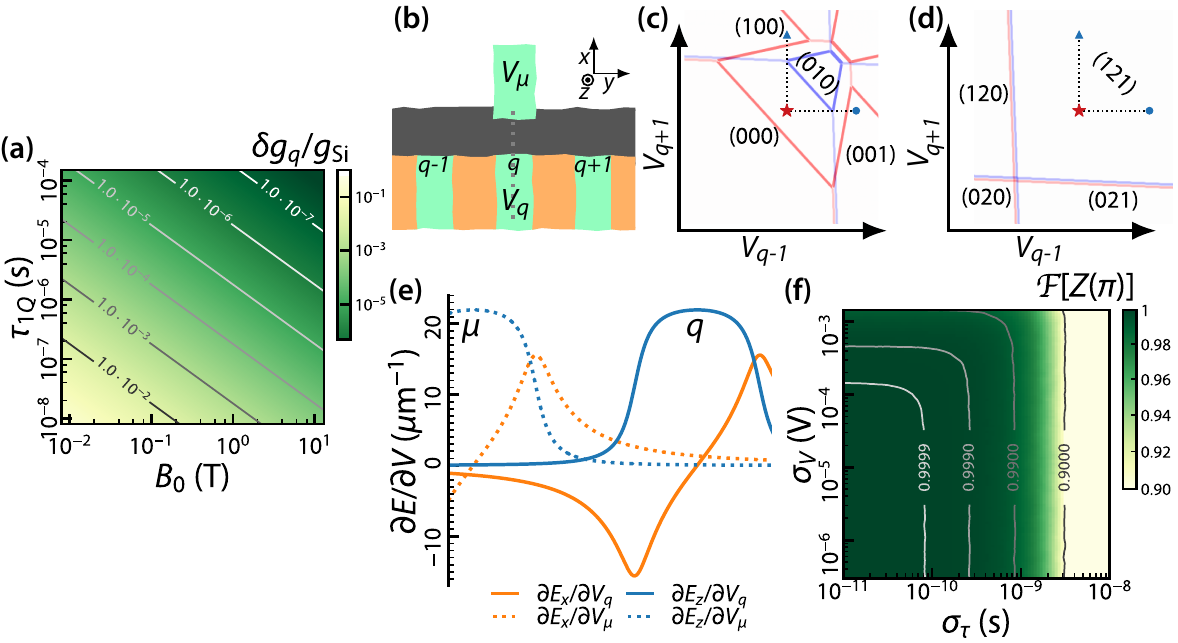}
\caption{
\label{fig:Z_rotation_gate}
\textbf{\textit{Z}-rotation gate using Stark shifts.} 
\textbf{(a)} Order of magnitude of Stark shift $\delta g_{q}$, with respect to the bulk value $g_{\mathrm{Si}} \approx 2.0$, as a function of external magnetic field $B_0$ and single-qubit gate time $\tau_{1Q}$, required for the $\pi$-rotation gate $Z(\pi)$. 
\textbf{(b)} Gate layout realisation of the $g$-factor tuning scheme with shuttling through quantum dots under gates $q-1$, to $q$, and to $q+1$. The voltage $V_{q}$ is used to tune the $g$-factor at site $q$, while the voltage $V_{\mu}$ is used to compensate for the change in the electrochemical potential due to the $g$-factor tuning.  
\textbf{(c)}-\textbf{(d)} Overlaid stability diagrams of the $(q-1,q,q+1)$ triple quantum dot  at the start and end (blue lines), and at the middle (red lines) of the shuttling sequence, illustrate the requirement for electrochemical potential compensation. 
Using a waveform as shown in Fig.~\ref{fig:nanogrid_layout} \textbf{(e)}, shuttling proceeds from the charge configuration $(n_{q+1}\,n_{q}\,n_{q-1}) = (001)$ (blue circle marker) to $(010)$ (red star marker), and to $(100)$ (blue triangle marker). 
\textbf{(c)} In the perfectly compensated case with $g$-factor tuning, the $(010)$ region opens up during the shuttling sequence. 
\textbf{(d)} In the non-compensated case, 
adjusting $V_{q}$ to tune the $g$-factor at $q$ 
causes the $(010)$ region to shift away from the ground state charge configurations.
\textbf{(e)} Electric field gradients evaluated along the cut shown as a gray dotted line in panel \textbf{(d)}. Electric field gradients due to $V_{q}$ are denoted as blue, and those due to $V_{\mu}$ as orange traces. 
\textbf{(f)} Estimated fidelity of $Z(\pi)$ as a function of variance in actual gate duration and voltage noise affecting $\delta g$, with fixed $\sigma_{G} = 10^{-3} g_{\mathrm{Si}}$, $B_{0} = 1$ T, and $\tau_{1Q} = 1\ \upmu$s.  
(see main text). 
}
\end{figure*}

The electron spin $g$-factor $g^{*}$, which defines the Larmor frequency $\omega_{0} = g^{*} \mu_{B} B_{0} / \hbar$ (where $\mu_B$ is the Bohr magneton and $B_0$ the applied dc magnetic field) can be shifted using electric fields from gate voltages. 
In SiMOS heterostructures the dominant Stark shift is understood to arise from wavefunction displacement with respect to the Si/SiO$_{2}$ interface~\cite{ruskov2018electron}. Linear or quasi-linear shifts $\delta g_{q} (V)$ with $\partial g/\partial E_{z}(V) = \pm (1\text{--}5) \times 10^{-3}\ (\mathrm{MV}/\mathrm{m})^{-1}$
have been reported for QDs in planar MOS devices
~\cite{veldhorst2014addressable,veldhorst2015spin,ferdous2018interface} with an in-plane magnetic field and $E_z$ applied perpendicular to the interface. The sign of the shift depends on the valley state of the electron~\cite{ruskov2018electron,veldhorst2015spin}. The roughness of the Si/SiO$_{2}$ interface also introduces some random contribution $G_{q}$ to the effective $g$-factor, which can be larger than the tuneability, $G_{q} = \pm \mathcal{O}(10^{-3} ... 10^{-2})$~\cite{ferdous2018interface}. Hence, the $g$-factor of a spin at some site $q$ can be considered as a combination of these shifts on some intrinsic value, $g_{\rm Si}$, such that $g^{*}_q = g_{\rm Si}+G_{q}+\delta g_{q}(V)$. In the following, we assume $g_{\rm Si} = 2.0$.
\par
We use the gate-voltage controllable $g$-factor Stark shift to perform relative single-qubit $Z(\varphi_{q}) = e^{- i \varphi_{q} \sigma_{z} / 2}$ gates by synchronously shuttling the qubit onto a QD structure (see columns $d+1$ and $d+3$ in Fig.~\ref{fig:nanogrid_layout} \textbf{(a)}) with predetermined dc gate voltages. Spins encoding the qubit state remain at the same site for a fixed time $\tau_{1Q}$ to acquire some phase (relative to a spin with $g=g_{\mathrm{Si}}$)
before synchronously shuttling forward. The gate $Z(\varphi_{q})$ is achieved by selecting a suitable g-factor shift 
\begin{align}
\delta g_{q}(V) &= \frac{ \varphi_q - 2 \pi r_q }{ \mu_B B_0 \tau_{1Q} \hbar^{-1} }. 
\label{eq:required_stark_shift}
\end{align}
Here, $r_q \in [ 0, 1)$ is selected from $G_q \mu_B B_0 \tau_{1Q} \hbar^{-1} := 2 \pi(n_q + r_q)$, for $n_q \in \mathbb{Z}$.
Since we only require phase matching up to $2 \pi$, the site-to-site randomness $G_q$ only contributes via $2 \pi r_q$, which remains in the same order of magnitude for any $G_{q}$. As a result, the randomness does not affect the required tuneability to attain a rotation angle over a target gate time.
Similarly, systematic phase shifts arising from the other QDs involved in shuttling can be accounted for as an effective contribution to $G_{q}$ and corrected in the same way. 
The order of magnitude of $\delta g_q(V)$ required to generate a $\pi$ phase shift at varying $\tau_{1Q}$ and $B_{0}$ is plotted in Fig.~\ref{fig:Z_rotation_gate} \textbf{(a)}, where we take $G_{q} = 0$ for simplicity. 
Tunabilities of at least $\delta g \approx \pm 3.6 \times 10^{-5}$ and $\delta g \approx \pm 3.6 \times 10^{-4}$ are required at $B_{0} = 1$ T for $\tau_{1Q} = 1\ \upmu$s and $\tau_{1Q} = 0.1\ \upmu$s, respectively.
\par
To realise such Stark shifts, we propose to employ the plunger gate as the $g$-factor tuning gate, and a neighbouring gate as a $\mu$-compensating gate (gate labelled with $\mu$ in Fig.~\ref{fig:Z_rotation_gate} \textbf{(b)}). 
This additional compensating gate is required to ensure the electrochemical potentials (see Supplementary Sec.~\ref{sec:electric_field_gradient_estimation}, Eqs.~\ref{eq:stark_shift_decomposition}-\ref{eq:electrochemical_potential_shift}), and hence QD electron occupancies remain correct under the applied electric field for $g$-factor tuning, as illustrated in the triple QD stability diagrams shown in Figs.~\ref{fig:Z_rotation_gate} \textbf{(c)}-\textbf{(d)}. 
In planar MOS, the plunger gate contributes dominantly to $E_z$, while the $\mu$-compensating gate E-field mostly to $E_x$, at site $q$, having negligible effect on the $g$-factor. In etched silicon devices, due to two facets of Si/SiO$_{2}$ interface (Fig.~\ref{fig:nanogrid_layout} \textbf{(f)}), we expect both split gates to contribute to the Stark shift, but as long as the effects of the two gates to the $g$-factor Stark shift and the electrochemical potential shift are asymmetric, the pair of gates allows to compensate for the shift in the electrochemical potential.
\par
Analytical estimates for the $E$-field derivatives $\partial E_z / \partial V_{j}$ and $\partial E_x / \partial V_{j}$ due to the plunger and $\mu$-compensating voltages $V_{q}$ and $V_{\mu}$, as a function of longitudinal position $x$, close to the MOS interface, are plotted in Fig.~\ref{fig:Z_rotation_gate} \textbf{(e)}, with simulation details given in Supplementary Sec.~\ref{sec:electric_field_gradient_estimation}. 
The estimated values of the $E_{z}$-gradient at site $q$, together with a conservative figure of $\partial g/\partial E_z = \pm 1 \times 10^{-3}\ (\mathrm{MV}/\mathrm{m})^{-1}$ suggest a required voltage shift $\mathrm{d}V_{q} = 615\ \mathrm{V} \times \delta g_{q}$. 
For example, a target Stark shift $\delta g_{\pi} = \pm 3.6 \times 10^{-5}$, would require the dc bias on the gate to be shifted $\mathrm{d}V_{q} = \pm 22$~mV.
The corresponding change in $\mu_q$ can be compensated with the $\mu$-compensating gate, by setting the voltage $\mathrm{d}V_{\mu} \approx - \alpha_{q q}/\alpha_{q \mu} \mathrm{d}V_{q}$, where $\alpha_{i j}$ is the lever arm between QD at site $i$ and gate $j$ 
(see Supplementary Sec.~\ref{sec:electric_field_gradient_estimation}).  
\par
We evaluate the sensitivity of this $Z(\varphi_{q})$ gate to noise using the so-called process fidelity~\cite{mayer2018quantum} between an ideal and a noisy unitary gate, for the case of $\tau_{1Q} = 1.0\ \upmu$s and $B_{0} = 1.0$ T.
We sample a value for $G_{q}$ from a normal distribution with width $\sigma_{G}$ set to $10^{-3} g_{\mathrm{Si}}$, and use it to find $\delta g_{q}$ (Eq.~\eqref{eq:required_stark_shift}) to hit $\varphi_{q} = \pi$ for an ideal gate.
For the noisy gate, we consider shuttling time errors of magnitude $\delta \tau$ that lead to gate time errors, and fluctuations in the gate voltage, $\delta V_{q}$, which lead to impacts in the $g$-factor according to $\delta g_{q} = \delta V_q / 615$ V. These noise contributions $\delta \tau$ and $\delta V_q$ are also sampled from normal distributions with varying widths of $\sigma_{\tau}$ and $\sigma_{V}$, respectively. 
\par
In Fig.~\ref{fig:Z_rotation_gate} \textbf{(f)}, we plot the resulting $Z(\pi)$ fidelities, averaged over $1000$ samples at each pair of noise levels. 
We find that the noise sources for the gate time and $g$-tuning errors add up independently. 
Noise levels of $\sigma_{\tau} \lesssim 0.08$ ns, i.e. $\sim 10^{-2} \tau_{\mathrm{s}}$ for the target shuttling time of $\tau_{\mathrm{s}} = 10$ ns, and $\sigma_{V} \lesssim 100\ \mu$V are required to achieve $\mathcal{F} \big[ Z(\varphi_{q}) \big] \geq 0.9999$. 
Charge noise acts equivalently to gate voltage fluctuations. Based on state-of-the art charge noise spectral densities in industrial SiMOS devices~\cite{elsayed2022low, spence2022probing, zwerver2022qubits}, 
we would expect voltage fluctuations of $\sigma_{V} \approx 30 ... 90$ $\mu$V over a bandwidth of $1\ \mathrm{Hz} - 1$ GHz, which is sufficient for our fidelity requirements, whereas there is less experimental data available on timing errors in shuttling.  
\subsection{Two-qubit gate family with gate-voltage-tuneable exchange strength}
\label{section:two_qubit_gate_family}

\begin{figure*}
\includegraphics[]{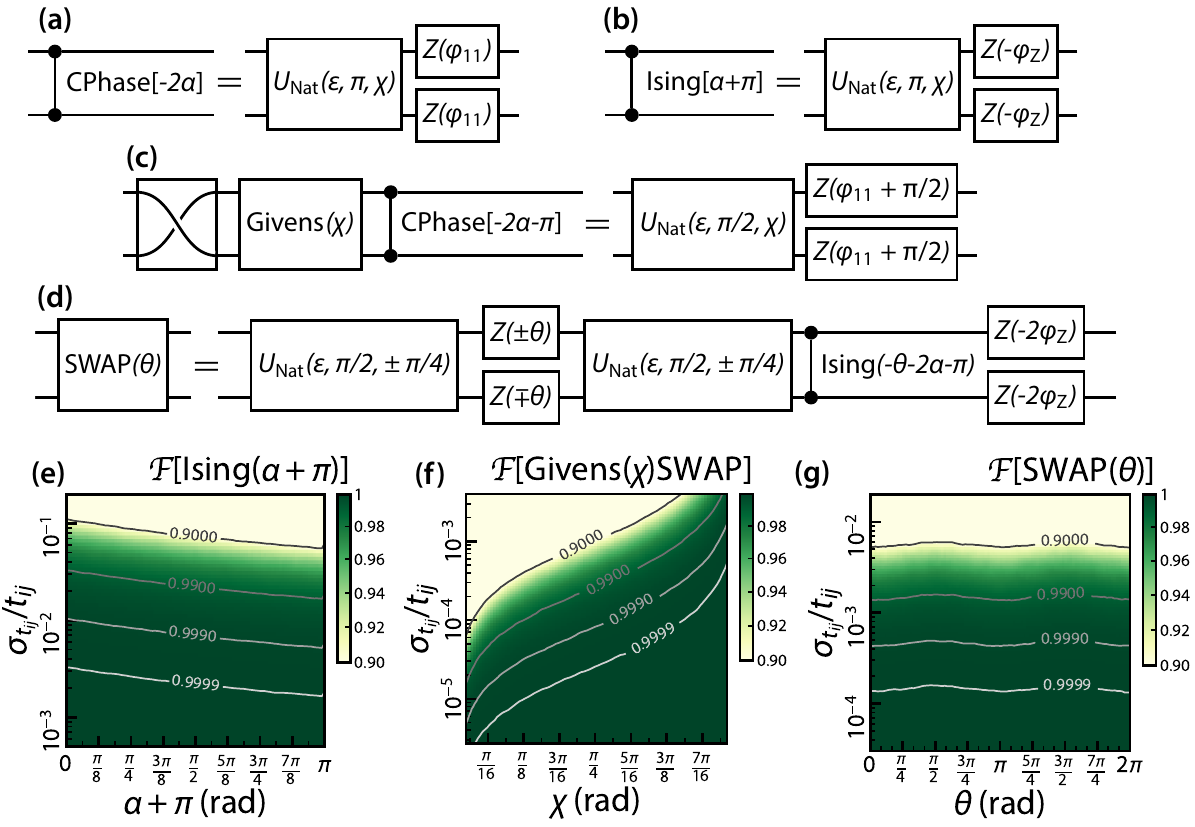}
\caption{
\label{fig:two_qubit_gate_family}
\textbf{Two-qubit gate family from nearest-neighbour exchange.}
\textbf{(a)}-\textbf{(d)} Circuit identities for the unitary time evolution $U_{\mathrm{NNE}}(\epsilon, \varphi, \chi)$ Eq.~\eqref{eq:unitary_nearest_neighbour_exchange_phi_chi_main}, describing nearest-neighbour exchange in the presence of Zeeman energy differences. Multiples of $2 \pi n$ are left out of the rotation angles for simplicity. 
\textbf{(a)}-\textbf{(b)} Choice of rotation angle $\varphi = \pi + 2 \pi n$ realises the phase gates \textbf{(a)} CPhase and \textbf{(b)} Ising $ZZ$-rotation gate. 
\textbf{(c)} Choice of $\varphi = \pi/2 + 2 \pi n$ realises a gate close to the Givens rotation, where the rotation angle $\chi$ depends on the ratio $\Delta E_{Z}/ J_{ij}$. 
\textbf{(d)} The SWAP-rotation gate can be constructed from the native unitary gate with $\varphi = \pi/2 + 2 \pi n$ and $\chi = \pi/4$, as two such native operations separated by single-qubit $Z$-rotation gates. The phases of the $m_{z} = \pm 1$ components are fixed by subsequent application of another phase gate and single-qubit $Z$-rotations.
\textbf{(e)}-\textbf{(g)} Fidelities for the native gates with \textbf{(e)} $\phi = \pi + 2 \pi n$,  realising the Ising $ZZ$-rotation gate, and \textbf{(f)} $\phi = \pi/2 + 2 \pi n$, realising the Givens$(\chi)$ SWAP operation, which, for $\chi = \pi/4$ is used in the composition of SWAP$(\theta)$ (see panel \textbf{(d)}), and \textbf{(g)} the composite SWAP$(\theta)$ rotation gate as a function of rotation angles and tunnel coupling variance $\sigma_{t_{ij}}/t_{ij}$. 
}
\end{figure*}

\par
To perform two-qubit operations, qubits are shuttled to neighbouring sites that connect adjacent pipes, in order to introduce an exchange interaction whose strength, $J_{ij}$, can be estimated by 
\begin{align}
J_{ij} 
&\approx
\frac{2 t_{ij}^2 \Delta K }{ 
\big[ (\Delta K)^2 - \epsilon(t)^2 \big] 
}.
\label{eq:perturbative_exchange_strength}
\end{align} 
Here, 
$\epsilon(t)$ is the detuning of the single-particle level spacings proportional to chemical potentials and $\Delta K$ is the difference between the on-site and inter-site charging energies. 
$J_{ij}$ can thus be in-situ modulated by the detuning using the plunger gates, or the tunnel coupling using the barrier gates. Both knobs can modulate the exchange strength over several order of magnitude. We choose to use module the tunnel coupling to allow us to operate in the centre of the $(1,1)$ charge configuration, where charge noise is minimised~\cite{reed2016reduced, martins2016noise}. 
\par
In the logic basis 
$\{ 
\ket{ \hspace*{-0.2em} \uparrow_{i} \, \uparrow_{j} }, 
\ket{ \hspace*{-0.2em} \uparrow_{i} \, \downarrow_{j} }, 
\ket{ \hspace*{-0.2em} \downarrow_{i} \, \uparrow_{j} }, 
\ket{ \hspace*{-0.2em} \downarrow_{i} \, \downarrow_{j} } 
\}$, 
the interaction between two exchange-coupled spins with a Zeeman energy difference $\Delta E_Z$ (see Supplementary Sec.~\ref{sec:nearest_neighbour_hamiltonian}, Eq.~\eqref{eq:two_site_one_orbital_exchange_Hamiltonian}) generates time-evolution
which is analogous to the single-qubit semiclassical Rabi dynamics in the $m_z = 0$ subspace, while the decoupled $m_z = \pm 1$ subspaces merely acquire phases according to their Larmor frequencies (see Supplementary Sec.~\ref{sec:nearest_neighbour_unitary}, Eq.~\eqref{eq:U_NNE_single_qubit_parameterisation}). In this analogy, within the $m_z = 0$ subspace, $\Delta_{ij} := \Delta E_{Z} + (J_{i} - J_{j})/2 \approx \Delta E_{Z}$ (also see Supplementary Sec.~\ref{sec:nearest_neighbour_hamiltonian}) corresponds to the qubit-drive detuning, $J_{ij}$ to the transversal coupling strength, and $\Omega_{ij} = \sqrt{ \Delta_{ij}^{2} + J_{ij}^{2} }/h$ to the Rabi frequency. 
\par
This time evolution is our native two-qubit operation. Figure~\ref{fig:two_qubit_gate_family} \textbf{(a)}-\textbf{(d)} illustrates some circuit identities obtained from it. To classify these operations, we may represent the unitary operation with two angle variables, as 
\begin{align}
&U_{\mathrm{Nat}} (\epsilon, \varphi, \chi) 
\, \hat{=} \,
e^{i \varphi \cos(\chi)} \times 
\nonumber \\
&
\begin{pmatrix}
e^{ -i \varphi_{Z} - i \alpha(\varphi, \chi)} \hspace*{-4.0em} & 0 & 0 & 0
\\
0 
&  
\hspace{-1.0em}
\big[ \hspace{-0.1em} \cos(\varphi) \hspace*{-0.1em} + i \hspace*{-0.1em} \sin(\chi) \sin(\varphi) \big]
\hspace*{-1.25em}
& 
-i \cos(\chi) \sin(\varphi)
& 
0
\\
0 
& 
-i \cos(\chi) \sin(\varphi) 
& 
\hspace*{-1.0em}  
\big[ \hspace{-0.1em} \cos(\varphi) \hspace*{-0.1em} - i \hspace*{-0.1em} \sin(\chi) \sin(\varphi) \big]
\hspace*{-0.5em}
& 
0
\vspace*{0.1em}\\
0 & 0 & 0 & \hspace*{-4.5em} e^{ i \varphi_{Z} - i \alpha(\varphi, \chi)}
\end{pmatrix} .
\label{eq:unitary_nearest_neighbour_exchange_phi_chi_main}
\end{align}
Here, the angle $\varphi = \tau_{2Q}/\Omega_{ij}$ is set by the gate time $\tau_{2Q}$ and the frequency $\Omega_{ij}$. 
The angle $\chi = \mathrm{arctan}(x)$ is set by the ratio $x = \Delta_{ij} / J_{ij}$. This ratio is related to the single-qubit analogue of the $B_{1}$ polar angle $\mathrm{arccos}(\Delta_{ij}/\Omega_{ij})$ via $\mathrm{arccos}(\Delta_{ij}/\Omega_{ij}) = \mathrm{arccot}(x)$. 
We also have $\varphi_{Z} = 
(E_{Z} + \Delta E_{Z}) \varphi / \Omega_{ij} $, and 
$\alpha(\varphi, \chi) = \varphi \cos(\chi)$. 
When $x \to 0$, coinciding with negligible Zeeman energy differences, the native operation~\eqref{eq:unitary_nearest_neighbour_exchange_phi_chi_main} reduces to the SWAP-rotation with the rotation angle given by $\varphi$ (viewed from the frame from which $E_{Z} = 0$). But even in the presence of $\Delta E_{Z}$, the native operation~\eqref{eq:unitary_nearest_neighbour_exchange_phi_chi_main} can be used to realise several familiar two-qubit operations. 
\par
One way to engineer desired gates starts by considering interaction times that correspond to particular numbers of completed rotations with respect to the single-qubit analogue of the Rabi frequency. In doing so, are left to fix $\chi$ to define the operation. Since we largely do not control the Zeeman energy difference, we choose the polar angle analogue with $J_{ij}$.
By choosing $\varphi = \pi + 2 \pi n$, we realise the diagonal two-qubit phase gates~\cite{meunier2011efficient}, as 
\begin{align}
&
Z (\varphi_{11})
\otimes 
Z (\varphi_{11})
e^{-i \alpha - i \varphi} 
U_{\mathrm{Nat}} (\epsilon \hspace{-0.15em} = \hspace{-0.15em} 0, \varphi = \pi \hspace{-0.15em} + \hspace{-0.15em} 2 \pi n, \chi)
=
\nonumber \\
&
\mathrm{CPhase}\big[ 
\hspace{-0.25em} - 2 \alpha(\varphi = \pi + 2 \pi n, \chi)
\big].
\label{eq:CPhase_from_Z1_Z2_phi_11_U_NNE_at_phi_equals_pi}
\end{align}
where we have defined $\varphi_{11} = - \varphi_{Z} - \alpha(\varphi, \chi)$, and
\begin{align}
&Z(-\varphi_{Z}) \otimes Z(-\varphi_{Z}) U_{\mathrm{Nat}} (\epsilon, \varphi =  \pi + 2 \pi n, \chi)
=
\nonumber \\
&\mathrm{Ising} 
\big[ 
\alpha( \varphi = \pi + 2 \pi n, \chi) + \pi
\big].
\label{eq:Ising_ZZ_rotation_from_Z1_Z2_phi_11_U_NNE_at_phi_equals_pi}
\end{align}
The circuits are visualised in Fig.~\ref{fig:two_qubit_gate_family} \textbf{(a)}-\textbf{(b)}. See Supplementary Sec.~\ref{sec:two_qubit_gates} for their matrix representations. 
These gates are maximally entangling (for certain rotation angles), but e.g. the SWAP gate, or any non-diagonal gate using just phase gates and single-qubit Z-rotations is not possible. In general, a larger set of gates allows more efficient decompositions for algorithms. For example, the variational eigensolvers for the Fermi-Hubbard model natively decompose into SWAP-rotations and single-qubit Z-rotations, so we show how to construct the SWAP-rotation from the native operation. 
\par
By instead choosing $\varphi = \pi/2 + 2 \pi n$, we obtain a gate close to the so-called Givens rotation (see Supplementary Sec.~\ref{sec:two_qubit_gates}, Eq.~\eqref{eq:Givens_rotation_gate}), as  
\begin{align}
&
Z(\varphi_{11} \hspace{-0.25em} + \hspace{-0.25em} \pi/2) 
\otimes 
Z(\varphi_{11} \hspace{-0.25em} + \hspace{-0.25em} \pi/2) 
e^{- i \alpha + i \varphi} \times
\nonumber \\
&U_{\mathrm{Nat}} (\epsilon, \varphi = \pi/2 + 2 \pi n, \chi) =
\nonumber \\
&\mathrm{CPhase}\big[ \hspace{-0.25em} - \hspace{-0.25em} 2 \alpha(\pi/2 + 2\pi n, \chi) + \pi \big] \mathrm{Givens}(\chi) \mathrm{SWAP}.
\label{eq:CPhase_Givens_SWAP_gate}
\end{align}
The circuit identity is illustrated in Fig.~\ref{fig:two_qubit_gate_family} \textbf{(c)}.
The angle of the Givens rotation is controllable with the polar angle analogue, although not all rotation angles are attainable equally easily. In particular, rotation angles of $0$ and $\pi/2$ would require negligible Zeeman energy difference or exchange strength, respectively. However, these cases are not interesting, since in the absence of Zeeman energy differences, we may employ a SWAP-rotation gate, and in the absence of an interaction the operation is non-entangling.
\par
The rotation angle $\chi = \pm \pi/4$ corresponds to the case where the (absolute value of the) Zeeman energy difference is equal to the exchange strength. Here, the $m_{z} = 0$ matrix elements simplify to $|U_{\mathrm{Nat}} (\epsilon, \pi/2, \pi/4)_{ij}| = 1/\sqrt{2}$. 
The native operation $U_{\mathrm{Nat}} (\epsilon, \pi/2, \pi/4)$ thus acts as a controlled Hadamard operation for the $m_{z} = 0$ subspace. The operation can be used to convert single-qubit $Z(\theta)$-rotations into SWAP-rotations, as
\par
\begin{align}
&\mathrm{SWAP}(\theta) =
\nonumber \\
&
- \hspace{-0.25em} e^{-i \theta \pm i \pi}
\,
Z \big( \hspace{-0.2em} - \hspace{-0.2em} 2 \varphi_{Z} \big) 
\hspace{-0.1em} \otimes \hspace{-0.1em}
Z \big( \hspace{-0.2em} - \hspace{-0.2em} 2 \varphi_{Z} \big) 
\mathrm{Ising} \big[ \hspace{-0.2em} - \hspace{-0.2em} \theta \hspace{-0.2em} - \hspace{-0.2em} 2 \alpha \hspace{-0.2em} - \hspace{-0.2em} 2 \varphi \big] \times \,
\nonumber \\
&
U_{\mathrm{Nat}}
\bigg( \hspace{-0.2em}
\epsilon, \frac{\pi}{2}, \pm \frac{\pi}{4}
\bigg)
\,
Z(\pm \theta) 
\hspace{-0.2em} \otimes \hspace{-0.2em} 
Z(\mp \theta) 
\, 
U_{\mathrm{Nat}}
\bigg( \hspace{-0.2em}
\epsilon, \frac{\pi}{2}, \pm \frac{\pi}{4}
\bigg).
\label{eq:SWAP_rotation_construction}
\end{align}
The circuit is illustrated in Fig.~\ref{fig:two_qubit_gate_family} \textbf{(d)}. 
We have written the circuit identity using the Ising gate for clarity, but in realising it using Eq.~\eqref{eq:Ising_ZZ_rotation_from_Z1_Z2_phi_11_U_NNE_at_phi_equals_pi}, we may absorb the $Z$-rotations into the final step, reducing physical gates from $6$ to $5$. 
It provides an exact method to perform SWAP-rotation operations, including the non-entangling SWAP gate, in the presence of Zeeman energy differences. The fidelity of this operation does not depend on the magnitude of $\Delta E_{Z}$ (as long as an equally large exchange strength is attainable), which means that the gate decomposition can be used with e.g. micromagnets, with which $\Delta E_{Z} = \mathcal{O}(1-10\ \mathrm{GHz})$~\cite{sigillito2019coherent}. 
\par
The strategy for choosing parameters for the $\varphi = \pi/2 + 2 \pi n$ operation is as follows (also see Supplementary Sec.~\ref{sec:native_operation_engineering} for further details). Knowing $\Delta E_{Z}$, we set $J_{ij} = |\Delta E_{Z}|$. We are required to set the gate time, as $\tau = (\pi/2 + 2 \pi n)/( \sqrt{2} J_{ij} )$. The gate time is limited by a minimum set by $\Delta E_{Z}$, and a resolution $2 \pi n/( \sqrt{2} J_{ij} )$, but we may use $g$-factor tuneability at QD $j$ for fine-tuning $\tau$ (after which $J_{ij}$ is recalculated). Smaller $g$-factor tuneability then requires a longer gate time to minimise gate time errors. 
\par
For example, setting the target gate time and rotation angle, as $\tau = 1\ \mu$s, and $\chi = \pi/4$, respectively, and assuming $\Delta K = 1$ meV, $\sigma_{G} = 10^{-3} g_{\mathrm{Si}}$, $B_{0}$ = 1 T yields the desired Givens-like gate with $x = \pm 1$, and with average exchange strength, number of rotations, and average timing errors of $J_{ij} \approx 32$ MHz, $n \approx 45$, and $\delta \tau_{2Q} \approx 0.1$ ns. For the phase gates, the protocol is similar, but $x$ is solved from the desired rotation angle. We note that typically, the two-qubit gates impose an independent restriction to the $g$-factor tuneability to ensure that fidelities are not limited by gate timing errors, which we find to be higher than the requirements for single-qubit gates. For example, in the above, we require approximately $\delta g \geq \pm 1 \times 10^{-4}$ (corresponding to $\mathrm{d} V_{q} \leq \pm 61$ mV). 
\par
The process fidelities of the native Ising$(\alpha + \pi)$, Givens$(\chi)$SWAP, and composite SWAP$(\theta)$ operations are shown in Fig.~\ref{fig:two_qubit_gate_family} \textbf{(e)}-\textbf{(g)}, as a function of rotation angle and relative variance in tunnel coupling noise, $\sigma_{t_{ij}}/ t_{ij}$. They are evaluated using the exact perturbative Hamiltonian (Supplementary Sec.~\ref{sec:nearest_neighbour_exchange},  Eq.~\eqref{eq:two_site_one_orbital_exchange_Hamiltonian}), at $\epsilon = 0$. We average over $N = 1000$ random $g$-factor pairs. For the Givens-like gate, we determine the sign of $\chi$ from the sign of the $g$-factor difference. All gates enable fidelities $\mathcal{F} \geq 0.9999$ for sufficiently low noise in the tunnel coupling. For example, charge noise in the barrier gate voltage propagates to noise in the tunnel coupling. The dependence of both rotation angles on tunnel coupling is reflected in the fidelities: angles that require higher $t_{ij}$ are more sensitive to tunnel coupling noise. However, since the rotation angles of the SWAP-rotation gate~\eqref{eq:SWAP_rotation_construction} arise from single-qubit $Z$-rotations, it's fidelity is approximately independent of rotation angles, and expected to be limited by the fidelity of the Givens$(\chi)$SWAP operation. 
\subsection{Transversal rotation gates} \label{section:transversal_qubit_control}
\par
A gate set enabling universal quantum computing requires another single-qubit gate besides the $Z(\varphi_{q})$ gate and a maximally entangling two-qubit gate~\cite{barenco1995elementary}.
To this end, we propose to realise a globally applied $\sqrt{X} = \textbf{I} (1 + i)/2 + \sigma_{x} (1 - i)/2$. 
The gate composes into a single-qubit rotation gate e.g. via the identity $Y(\varphi_{q}) := \sqrt{X} Z(\varphi_{q}) \sqrt{X}^{\dagger}$. 
\par
In the nanogrid, a $\sqrt{X}$ gate can be applied globally by providing a small $B_{1}$ perpendicular to $B_{0}$ from large resonant structures, such as a dielectric 3D cavity resonator~\cite{vahapoglu2022coherent}, or a superconducting resonator based on a coplanar waveguide patterned e.g. over the metal gate layers~\cite{rausch2018superconducting}, with a resonance frequency coinciding with the average qubit frequency $f_{\mathrm{Si}} \approx 28$ GHz, and a quality factor $Q \approx 100$ to cover a bandwidth of $280$ MHz corresponding to $G_{q} \leq \pm 10^{-2}$ ($ > 5 \sigma_G$, assuming $\sigma_G = 10^{-3} g_{\mathrm{Si}}$).  
Global control allows avoiding issues related to crosstalk and impedance matching, which would be a challenge with partially or fully local broadband structures~\cite{dehollain2012nanoscale}. 
\par
The global $X$-control together with pipelining provides an extra limitation for the circuit compilation and density of pipelining. Since all qubits on the pipeline undergo the $\sqrt{X}$ pulses, the algorithms must be compiled with periodic $X$-control. A simple example code block for the pipelined, global-$X$-controlled compilation is given by
\begin{align}
[ Z(\theta_{1}), X(\pi/2), Z(\theta_{2}), Z(\theta_{3}), X(\pi/2), \mathrm{Native}(\theta_{4}) ] \times D/6,
\label{eq:example_code_block}
\end{align} 
where $\mathrm{Native}(\theta_{4})$ is a native two-qubit gate of the system, as discussed in Sec.~\ref{section:two_qubit_gate_family}. This code block allows pipelining at a filling density of one in every three (logical) columns. The code block~\eqref{eq:example_code_block} maps to an equal density of single-qubit $Z$-rotation gates, $Y$-rotation gates, and native two-qubit interactions.  
Due to the variations in g-factor discussed above, driving all spins with a single drive tone is challenging. 
We instead opt for multitone driving and frequency binning~\cite{fogarty2022silicon}, which is discussed in Supplementary Sec.~\ref{section:frequency_binning}. We show that reaching $\mathcal{F} \geq 0.9999$ requires $\sigma_{B_{1}} < 0.1\ \mu$T and $\sigma_{\tau} < 0.2$ ns.
\section{Application as bespoke hardware for a NISQ eigensolver}
\label{section:application_as_bespoke_NISQ_hardware}
\begin{table*}
\renewcommand{\arraystretch}{2}
\setlength{\tabcolsep}{1ex}
\begin{tabular}{lclll}
\toprule
\textbf{Operation}
&
\textbf{Symbol}
&
\textbf{Time}
&
\textbf{Method}
\\ \midrule
\makecell[l]{\textbf{Shuttling}}
&
$\tau_{s}$
&
$10$ ns
&
Global bucket-brigade pulses
\\ 
\makecell[l]{\textbf{Single-qubit $Z$-rotations}}
&
$\tau_{1Q}$
&
$1\ \upmu$s 
 ($0.1\ \upmu$s)
&
Local $g$-factor Stark shifts
\\ 
\makecell[l]{\textbf{Single-qubit $X$-rotations}}
&
$\tau_{1X}$
&
$3 \tau_{1Q} + 8 \tau_{s} = 3.06\ \upmu$s
 ($0.36\ \upmu$s)
&
Global $B_{1}$-drive frequency bins
\\ 
\makecell[l]{\textbf{Two-qubit native gates}}
&
$\tau_{2Q}$
&
$1\ \upmu$s 
 ($0.1\ \upmu$s)
&
Native exchange interaction
\\ 
\makecell[l]{\textbf{Two-qubit phase gates}}
&
$\tau_{2P}$
&
$\tau_{2Q} + \tau_{1Q} + 3\tau_{s}$ 
$= 2.0\ \upmu$s
 ($0.23\ \upmu$s)
&
\makecell[l]{Composite from native gates
\\
and $Z$-rotations}
\\ 
\makecell[l]{\textbf{Two-qubit SWAP-rotations}}
&
$\tau_{2S}$
&
\makecell[l]{$3\tau_{2Q} + 2\tau_{1Q} + 12\tau_{s}$ 
$= 5.1\ \upmu$s 
}
 ($0.62\ \upmu$s)
&
\makecell[l]{Composite from native gates
\\ and $Z$-rotations}
\\ \bottomrule
\end{tabular}
\caption{
\label{tab:pipeline_hardware_protocols}
\textbf{Qubit control protocols on the pipeline.}
Summary of the protocols introduced in Section~\ref{section:implementation_with_nanogrid}. We have marked the gate times assuming $\tau_{1Q} = \tau_{2Q} = 0.1\ \upmu$s in parenthesis. Numbers that are not in parentheses assume $\tau_{1Q} = \tau_{2Q} = 1\ \upmu$s. Faster gate times require higher attainable $g$-factor tuning to prevent gate time errors from dominating the infidelities. 
}
\end{table*}
\begin{table*}
\renewcommand{\arraystretch}{1.8}
\setlength{\tabcolsep}{1.5ex}
\begin{tabular}{lcll}
\toprule
\textbf{Eigensolver flow level}
&
\textbf{Symbol}
&
\textbf{Run-time overhead} 
&
\textbf{Pipelined run-time overhead}
\\ \midrule
\textbf{Run-time per circuit configuration}
&
$\tau_{\mathrm{config}} $
&
\makecell[l]{
$\big[ D_{1Q} \big( \tau_{1Q} + 3 \tau_{s} \big)$
\\
$+ D_{2Q} \big( \tau_{2S} + 3 \tau_{s} \big) \big] N_{\mathrm{reps}}$
\\
$= 26$ minutes ($3.8$ minutes)
}
&
\makecell[l]{
$\big[ \big(D_{1Q} + 2 N_{\mathrm{reps}} \big) \big( \tau_{1Q} + 3 \tau_{s} \big) 
$
\\
$+ \big(D_{2Q} + 2 N_{\mathrm{reps}} \big) \big( \tau_{2S} + 3 \tau_{s} \big) \big]$
\\
$ = 1.7$ seconds ($0.25$ seconds)
}
\\ 
\textbf{Circuit configurations}
&
$N_{\mathrm{configs}}$
&
$3900$
&
$3900$
\\ 
\textbf{Iterations}
&
$N_{\mathrm{iters}}$
&
$100$
&
$100$
\\ 

\makecell[l]{
\textbf{Total run-time}\\
}
&
$\tau_{\mathrm{run}}$
&
\makecell[l]{
$N_{\mathrm{iters}} N_{\mathrm{configs}} \tau_{\mathrm{config}}$
\\
$= 230$ months ($35$ months)
}
&
\makecell[l]{
$N_{\mathrm{iters}} N_{\mathrm{configs}} \tau_{\mathrm{config}}$
\\
$= 7.9$ days ($11$ hours)
}
\\ \bottomrule
\end{tabular}
\caption{
\label{tab:application_to_NISQ_eigensolver}
\textbf{NISQ Fermi-Hubbard eigensolver run-time on the pipeline.}
The times in parentheses assume $\tau_{1Q} = \tau_{2Q} = 0.1\ \upmu$s, whereas otherwise we assume, that $\tau_{1Q} = \tau_{2Q} = 1\ \upmu$s. 
}
\end{table*}
We summarise the proposed qubit control protocols for operating the silicon spin qubit pipeline from Sec.~\ref{section:implementation_with_nanogrid} in Table~\ref{tab:pipeline_hardware_protocols}. The requirements of synchronous shuttling, lack of local runtime control, and qubit frequency variability leads to protocols where we fix gate times and adjust qubit frequencies with dc voltage tuneable parameters, namely the $g$-factor (qubit frequency) and the exchange strength. Each of these protocols are feasible up to fidelities $\mathcal{F} \geq 0.9999$ in the presence of noise, in the realistic scenario where qubit frequency variabilities are larger than the frequency tuneability. 
\par
We now exemplify how these elements propagate into solving a quantum computing problem in the NISQ era. Here, we focus on the variational eigensolver for the Fermi-Hubbard model, where the resources required to run the algorithm on a set of physical qubits have been estimated in~\cite{cai2020resource}, and where the task is to estimate the ground state of a $5 \times 5$ Fermi-Hubbard Hamiltonian. See supplementary Sec.~\ref{sec:variational_quantum_eigensolvers} for details of this algorithm. At the low level, the algorithm breaks down into SWAP-, and single-qubit $Z$-rotation gates for the so-called (simulated) state initialisation and (simulated) state evolution stages, while the bit-string (physical qubit) initialisation and readout stages also require qubit-selective qubit-flip $X(\pi)$ and basis-change $X(\pi/2)$ operations. We assume that we are not limited by initialisation or readout times. The single-qubit gate times, and the two-qubit gate time errors are upper limited by the qubit frequency tuneability. While the tuneability has not been studied on a large number of devices, based on the literature, we expect the processor clockspeeds no slower than $1$ MHz (at $B_{0} = 1$ T). This means that to run an algorithm of depth $10000$, for example, we require qubit $T^{*}_{2} \geq 10$ ms. 
\par
We summarise the estimated pipelined run-time, and contributions to the run-time, of the Fermi-Hubbard variational eigensolver in Table~\ref{tab:application_to_NISQ_eigensolver}, using the resource estimates from~\cite{cai2020resource}.
There are several layers at the high level. The algorithm requires a number of \textit{iterations}. Each iteration consists of a number of \textit{runs}, which is equal to the number of circuit configurations multiplied by runs per configuration. The number of \textit{circuit configurations}, in turn, is determined by the number of parameters, number of measured observables, and the number of noise levels, which is part of an error mitigation protocol~\cite{cai2021multi}. The number of \textit{runs per configuration} is determined by the number of runs for one parameter, to measure one set of commuting observables, and the extra sampling cost for error mitigation. For each parameter-observable-set-specific run, we may then evaluate the circuit run time. 
\par
Two types of classical parallelisation are possible. As discussed in Sec.~\ref{section:qubit_pipeline}, pipelining allows to perform $N_{r}$ number of repetitions for a circuit of depth $D$ in time $(D + N_{r})\tau$. For example, at maximal filling, this runtime scaling is $(D + 2 N_{r})\tau$. 
For single-qubit and two-qubit gate depths of $D_{1Q}$ and $D_{2Q}$, the run-time for $N_{r}$ repetitions on the nanogrid pipeline (Fig.~\ref{fig:nanogrid_layout} \textbf{(a)}) with three physical shuttling steps between gates is then $(D_{1Q} + 2 N_{r}) (\tau_{1Q} + 3 \tau_{s}) + (D_{2Q} + 2 N_{r}) (\tau_{2S} + 3 \tau_s)$. For $D_{1Q} = 1174$ and $D_{2Q} = 2196$~\cite{cai2020resource}, and the gate time estimates for $Z$-rotation and SWAP-rotation gates from Table~\ref{tab:pipeline_hardware_protocols}, we find that pipelining reduces the run-time per circuit configuration from $25.5$ minutes (assuming $\tau_{1Q} = 1\ \upmu$s) to $1.74$ seconds, i.e. by roughly a factor of $880$.  
\par
We may also run e.g. different circuit configurations on different physical pipeline processors in parallel. 
To estimate the footprint of the pipeline processor, we expect the width of a single pipe to be approximately $340$ nm, with a same-layer gate pitch of $100$ nm. Likewise, we expect the length of a single-qubit or two-qubit gate step to be approximately $190$ nm. Then, for $N = 25$ qubits, and a circuit depth of $D = D_{1} + D_{2} = 1174 + 2196 = 3370$ quantum logic gates, we estimate the footprint of a single pipeline qubit processor to be $8.5\ \mu$m $\times 640.3\ \mu$m. 
\section{Conclusions and discussion} \label{section:Conclusion}
We have analysed a qubit pipeline architecture for realizing gate based quantum computation in the NISQ era. The architecture minimizes run-time local control resources, utilizing instead global run-time, and local pre-configuration control.  
This is made possible by a combination of an increased qubit grid layout size, and by synchronized operation, where steps of qubit state shuttling and quantum logic gates alternate. 
\par
Having described the architectural paradigm, we focused on a physical implementation case-study in the SiMOS electron spin qubit platform. Here, we laid out qubit control protocols under the pipeline- and platform-specific restrictions, demonstrating each theoretically with NISQ-high fidelities while remaining robust against qubit frequency variabilities characteristic for the platform. Our main focus has been to address this frequency variability, while we may improve robustness against noise with bespoke control methods in the future. Most of the elements are possible to implement with present-day technology without further advances, but we expect more microwave engineering efforts to designing and testing switchable, dense transmission lines or resonators, and their ability to support a finite frequency bin. We then assessed the performance of this architecture for a NISQ variational eigensolver task. In the future, it may be possible to decrease the number of required gates per run by more directly utilising the native two-qubit gate family that arises from nearest neighbour exchange in the presence of Zeeman energy differences. 
The silicon spin qubit platform is well-suited to the pipeline approach, but the concept may also be implemented in other architectures, such as with trapped ions, or with superconducting qubits by replacing shuttling with SWAPs. 
\par
Indeed it would be an interesting topic of further work to explore an implementation based on SWAPs. There, the qubit grid remains fully occupied and stationary, and useful quantum information is transferred forward through the array (while states carrying no quantum information propagate in reverse). As before, the horizontal density of quantum information in the array can be adjusted as required (e.g. to accommodate initialisation and measurement times) by introducing buffer states which do not participate in the calculation. Back-propagating states or buffer states do not interfere with the calculation 
due to the non-entangling nature of the SWAP gate. 
\section{Acknowledgements}
Balint Koczor is acknowledged for a useful discussion regarding gate fidelities. 
SMP acknowledges the Engineering and Physical Sciences Research Council (EPSRC) through the Centre for Doctoral Training in Delivering Quantum Technologies (EP/L015242/1). MFGZ acknowledges support from UKRI Future Leaders Fellowship [grant number MR/V023284/1].  
\newpage
\bibliography{references/references.bib}
\end{document}


\title{Supplement for Pipeline quantum processor architecture for silicon spin qubits}
\newcommand\QMT{\,Quantum Motion, 9 Sterling Way, London N7 9HJ, United Kingdom}
\newcommand\UCL{\,London Centre for Nanotechnology, University College London, London WC1H 0AH, United Kingdom}
\newcommand\Oxford{\,Department of Materials, University of Oxford, Parks Road, Oxford OX1 3PH, United Kingdom}
\author{S. M. Patom{\"a}ki}
\affiliation{\QMT}
\affiliation{\UCL}
\email{ucapsmp@ucl.ac.uk}
\author{M. F. Gonzalez-Zalba} 
\affiliation{\QMT}
\author{M. A. Fogarty} 
\affiliation{\QMT}
\author{Z. Cai} 
\affiliation{\QMT}
\affiliation{\Oxford}
\author{S. C. Benjamin}
\affiliation{\QMT}
\affiliation{\Oxford}
\author{John J. L. Morton}
\affiliation{\QMT}
\affiliation{\UCL}
\date{\today}
\maketitle
\tableofcontents
\appendix
\setcounter{section}{18}
\renewcommand{\thesection}{\Alph{section}}
\renewcommand\thefigure{\thesection\arabic{figure}}
\section{Supplementary information}
\subsection{Programmable pixelgrid} 
\label{sec:pixelgrid}
We obtain an algorithm decomposition-reconfigurable pipeline using a dense 2D array of quantum dot defining metal gates. We outline this so-called pixelgrid in Fig.~\ref{fig:pixelgrid}. Gate voltages determine which sites act as pipes, and which stages of the algorithm contain two-qubit gates. The design has the advantage of optimising gate decompositions, but the disadvantages associated with maximal density, such as crosstalk, and highly specialised demands for the fabrication of routing. Such routing demands have thus far only been demonstrated for $2 \times N$ QD arrays~\cite{ha2021flexibleSupplement,weinstein2023universalSupplement}.   
\par
\documentclass[../../../main/main.tex]{subfiles}
\begin{document}
\begin{figure}
\includegraphics[]{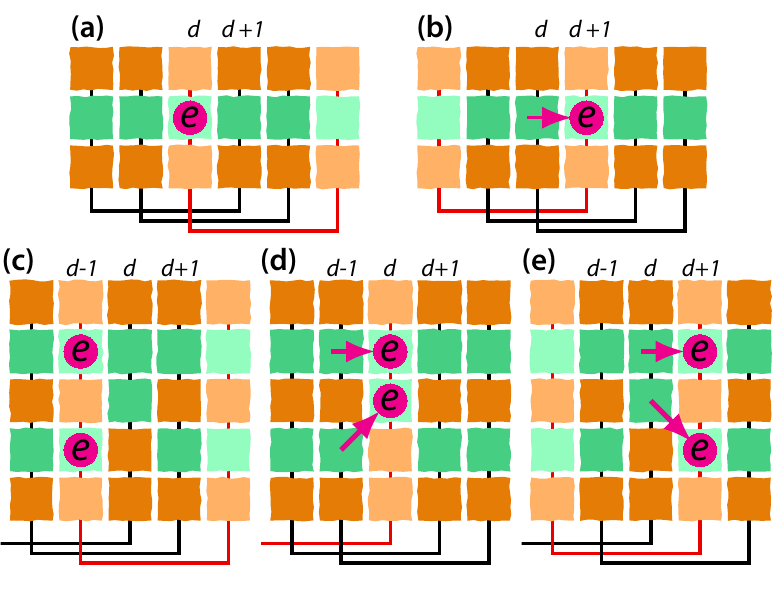}
\caption{
\label{fig:pixelgrid}
\textbf{Programmable pixelgrid.} 
A dense quantum dot array provides a reconfigurable implementation of the pipeline. Squares, or pixels, represent quantum dot defining metal gates. The decomposition of the algorithm to single-, and two-qubit gates determines which gates acts as plungers (green) and which as barriers (orange). 
\textbf{(a)}-\textbf{(b)} Shuttling forward along the pixelgrid is implemented with three-stage shuttling. 
\textbf{(c)}-\textbf{(e)} Two-qubit interactions between pipes are implemented with diagonal shuttling. 
}
\end{figure}
\end{document}
\subsection{Shuttling}
\subsubsection{Minimizing shuttling errors}
\label{sec:minimizing_shuttling_errors}
As discussed in the main text, non-adiabatic time evolution can create shuttling errors via so-called Landau-Zener transitions. 
Thus, high-fidelity shuttling should be performed adiabatically with respect to inter-dot charge transition tunnel rates, and using ramp rates which are not resonant with valley-orbit transitions. 
\par
For a simple ballpark figure of adiabaticity, we may estimate probability for charge state error by evaluating the first-order, or single-passage, Lanzau-Zener transition probability, which is given by $P_{\mathrm{LZ}}~=~\exp\big(-2 \pi \delta \big)$, where $\delta = t_{ij}^2/(4 \nu)$, and furthermore $\nu = A \hbar \omega$ is the approximate drive velocity for a ramp with amplitude $A$ (in units of energy) and angular frequency $\omega$~\cite{shevchenko2010landauSupplement}. 
Over a typical voltage range across an inter-dot-charge transition, $A = e \alpha \times 25$ mV, where $\alpha = 0.1$ is a typical lever arm. 
At tunnel coupling $t_{ij}/h = 20$ GHz, the ramp can be performed adiabatically ($P_{\mathrm{LZ}} < 10^{-4}$) with shuttling times of $9.1$ ns or more (corresponding to $\omega/(2\pi) \lesssim 110$ MHz). 
\subsubsection{Control line footprints}
\label{sec:control_line_footprints}
A bias tee where current is not expected to flow can be realised using a resistor and a capacitor. Assuming a sheet resistance of $\rho = 100\ \Omega /$square~\cite{jin2020modelingSupplement}, the footprint of a single $R_{\mathrm{tee}} = 10\ \mathrm{k}\Omega$ resistor would be $50\ \mathrm{nm} \times 5\ \upmu \mathrm{m}$. Likewise, assuming a capacitance per area of $1$ pF/$\upmu$m$^{2}$~\cite{boter2022spiderwebSupplement}, a $C = 159$ pF capacitor, to hit an $RC$-constant cutoff frequency of $f_{\mathrm{cutoff}} = 100$ kHz, would take around $159\ \mu$m$^{2} \approx 12.6\ \upmu\mathrm{m} \times 12.6\ \upmu\mathrm{m}$. For a single column of $50$ qubits, bias tees could be fitted into an area of approximately $630\ \upmu$m $\times 12.7$ $\mu$m. The $1$-to-$N$ power splitter can be realised as e.g. a common-source $N$-parallel-MOSFET. 
\subsection{Electric field dependence on applied gate voltages}
\label{sec:electric_field_gradient_estimation}
For the $g$-factor Stark shift
\begin{align}
\delta g_{q}(V_{j}) &= \frac{ \partial g }{ \partial E_i } \frac{ \partial E_i }{ \partial V_j } \mathrm{d}V_j,
\label{eq:stark_shift_decomposition}
\end{align}
where $E_i$ are the electric field components and $V_j$ the associated gate voltages, we have a corresponding change in electrochemical potential of 
\begin{align}
\mathrm{d}\mu_{q} = -e \sum_{j} \alpha_{qj} \, \mathrm{d}V_{j},
\label{eq:electrochemical_potential_shift}
\end{align} 
where $\alpha_{qj}$ is the lever arm from dot $q$ to gate $j$. The change in $\mu_q$ would be detrimental to the shuttling scheme when uncompensated, which can be illustrated with the triple QD stability diagrams, shown in main text Figs.~\ref{fig:Z_rotation_gate} \textbf{(e)}-\textbf{(g)}. 
Stability diagrams show the boundaries of regions of constant charge as a function of two (or more, in higher-dimensional graphs) gate voltages. 
In a successful shuttling sequence, the shuttling waveform takes the electron from the charge configuration $(n_{q+1}\,n_{q}\, n_{q-1}) = (001)$, to $(010)$, and to $(100)$, as illustrated in Fig.~\ref{fig:Z_rotation_gate} \textbf{(e)}. Uncompensated $g$-factor modulation using $V_{q}$ can lead to a change in the proximal ground state charge configurations, 
which under globally applied shuttling sequence would lead to an error in the charge state as illustrated in Fig.~\ref{fig:Z_rotation_gate} \textbf{(f)}. 
\par
To estimate the relative contributions of the plunger gate and the $\mu$-compensating gate, we solve for the derivatives $\partial E_i / \partial V$ as follows. We model a metal gate as a rectangular infinitely thin charge sheet centered at origin. The sheet has width $a$, length $b$, and a uniform charge density $\sigma$. The electric field components can be expressed as the double integrals of Eqs.~\eqref{eq:Ex_integral}-\eqref{eq:Ez_integral},
\begin{figure*}
\begin{align}
E_x &=
k_e \sigma \int_{-a/2}^{a/2} \mathrm{d}x' (x - x') 
\int_{-b/2}^{b/2} \mathrm{d}y'  \big[ (x - x')^{2} + (y - y')^{2} + z^{2} \big]^{-3/2}, 
\label{eq:Ex_integral}
\\
E_y &=
k_e \sigma 
\int_{-a/2}^{a/2} \mathrm{d}x' \int_{-b/2}^{b/2} \mathrm{d}y' (y - y') \big[ (x - x')^{2} + (y - y')^{2} + z^{2} \big]^{-3/2}, 
\label{eq:Ey_integral}
\\
E_z &= 
k_e \sigma z 
\int_{-a/2}^{a/2} \mathrm{d}x' \int_{-b/2}^{b/2} \mathrm{d}y' \big[ (x - x')^{2} + (y - y')^{2} + z^{2} \big]^{-3/2}, 
\label{eq:Ez_integral}
\end{align} 
\end{figure*}
where $k_e = 1/(4 \pi \epsilon_0 \epsilon_r)$. 
These can be computed with Mathematica
\footnote{
Mathematica 
$12.1.1.0$ 
is able to integrate the planar components $E_{x}$ and $E_{y}$, and the first integral of $E_{z}$. The second integral is evaluated after a transform of variables.
}, 
or by hand using substitutions, which for the second integral of~\eqref{eq:Ez_integral} would read: $x'' = x - x'$, where $\mathrm{d}x'' = \mathrm{d}x'$, followed by the trigonometric substitution $x''/z = \mathrm{tan}(\theta)$, with $\mathrm{d}x'' = \mathrm{d} \theta \cos(\theta)^{-2}$. The substitution allows to simplify the integrand in a form which is integrable using trigonometric identities. The resulting functions can be expressed in terms of elementary functions. 
\par
Once we fix a heterostructure of planar dielectrics with large surface areas, we obtain a relationship between applied gate voltage $V$ and the effective charge density $\sigma$, as
\begin{align}
V &= - \sum_{i} \int_{z_{i}}^{z_{i+1}} \mathrm{d} \textbf{l} \cdot \textbf{E}(x,y,z), \label{eq:V_sigma_relationship}
\end{align}
where the sum is taken over the interfaces between different materials from the charge sheet to the ground plane. 
For example, 
there is a single layer of dielectric and substrate between the gate and the ground plane in a planar MOS structure such that $z_0 = 0$, $z_1 = d_{\mathrm{ox}}$, and $z_{2} = d_{\mathrm{ox}} + d_{\mathrm{Sisub}}$, where $d_{\mathrm{ox}}$ and $d_{\mathrm{Sisub}}$ are widths of the dielectric and the Si substrate, respectively. 
For a path perpendicular to the interfaces 
 $\mathrm{d} \textbf{l} \cdot \textbf{E} = \mathrm{d}z \, E_z$. The result is a linear relationship 
\begin{align}
\sigma = \sigma(V) = a_{\sigma} V, \label{eq:sigma_V_relationship}
\end{align} where $a_{\sigma}$ is a constant with respect to the coordinates, and is a function of the geometry. 
The relation~\eqref{eq:sigma_V_relationship} allows us to express $\textbf{E}(\sigma) = \textbf{E}(\sigma(V))$, which allows us to evaluate the derivatives $\partial E_i / \partial V$ analytically. Since $\textbf{E} \propto \sigma \propto V$, $\partial E_i / \partial V \propto \textbf{E}$, and $\partial E_i / \partial V$ is independent of $V$. Table~\ref{tab:parameters_for_electric_field_gradient_estimation} summarises the parameters used in this simulation.
\par
\begin{table}
\begin{tabular}{llll}
\toprule
\textbf{Variable} & \textbf{Symbol} & \textbf{Value} & \textbf{Unit}
\\ \midrule
QD length along $x$ & $w_{x}$ & $50$ & nm
\\ \midrule
QD length along $y$ & $w_{y}$ & $50$ & nm
\\ \midrule
Si/SiO$_{2}$ oxide thickness & $d_{\mathrm{ox}}$ & $5$ & nm
\\ \midrule
Gate to gate distance & $\Delta_{\mathrm{\mu q}}$ & $40$ & nm
\\ \midrule
Si substrate thickness & $d_{\mathrm{Sisub}}$ & $0.5$ & mm
\\ \midrule
Si dielectric constant & $\epsilon_{\mathrm{Si}}$ & $11.8$ & 
\\
SiO$_{2}$ dielectric constant & $\epsilon_{\mathrm{ox}}$ & $3.8$ & 
\\ \bottomrule
\end{tabular}
\caption{
\label{tab:parameters_for_electric_field_gradient_estimation}
\textbf{Simulation parameters for the electric field gradient simulations.}
}
\end{table}
\par
For planar MOS or mostly $\pm$ z-valley-lying wavefunctions, the effect of $E_z$ due to $V_{q}$ dominates. This is because 
$\partial E_x / \partial V_{q} \approx 0.010 \, \partial E_z / \partial V_{q}$,  
$\partial E_x / \partial V_{\mu} \approx 0.062 \, \partial E_z / \partial V_{q}$, and 
$\partial E_z / \partial V_{\mu} \approx 0.0037 \, \partial E_z / \partial V_{q}$
at site $q$ (see Fig.~\ref{fig:Z_rotation_gate} \textbf{(g)}). 
We also expect $\partial g / \partial E_z \gg \partial g / \partial E_x$. More generally, when both gates Stark shift the $g$-factor, compensation is possible, as long as the effects of the gates to $g$-tuning and $\mu_{q}$ are asymmetric. 
Then  Eq.~\eqref{eq:stark_shift_decomposition} simplifies to $\delta g_{q} (V) \approx \big( \partial g / \partial E_{z} \big) \, \big( \partial E_{z} / \partial V_{q} \big) \, \mathrm{d} V_{q}$. 
\par
As discussed in the main text, tuneability of $\delta_{g \pi} = \pm 3.6 \times 10^{-5}$ may require plunger gate voltage shifts of $\mathrm{d}V_{q} = \pm 0.022$ V, which can be compensated with $\mathrm{d}V_{\mu} \approx - \alpha_{q q}/\alpha_{q \mu} \mathrm{d}V_{q}$. 
In addition, perfect compensation requires $\alpha_{qq}/\alpha_{q\pm1 \, q} = \alpha_{q \mu}/\alpha_{q\pm1\, \mu}$, where $\alpha_{q\pm1 \, q}$ and $\alpha_{q\pm1 \, \mu}$ are the lever arms to of the subsequent and prior QDs to the plunger gate of $q$ and the $\mu$-compensating gate, respectively. The stability diagrams in Fig.~\ref{fig:Z_rotation_gate} \textbf{(e)} are in fact simulated using the $\mu$-compensation scheme described above. The results are identical to those in the absence of $g$-factor tuning. The stability diagram is simulated from the ground state energy of the electrostatic Hamiltonian. See e.g.~\cite{patomaki2023elongated} for details of the simulation. 
\subsection{Nearest-neighbour exchange} \label{sec:nearest_neighbour_exchange}
\subsubsection{Nearest-neighbour Hamiltonian}
\label{sec:nearest_neighbour_hamiltonian}
The two-site (for sites $i$, $j$), one-orbital Fermi-Hubbard Hamiltonian is block-diagonalised using a second order Schrieffer-Wolff transformation, separating $(1,1)$ as low-energy states compared to $(2,0)$ and $(0,2)$ states~\cite{meunier2011efficientSupplement}. 
The resulting block-diagonalised Hamiltonian for the $(1,1)$ subspace, in the basis 
$\{ 
\ket{ \hspace*{-0.2em} \uparrow_{i} \, \uparrow_{j} }, 
\ket{ \hspace*{-0.2em} \uparrow_{i} \, \downarrow_{j} }, 
\ket{ \hspace*{-0.2em} \downarrow_{i} \, \uparrow_{j} }, 
\ket{ \hspace*{-0.2em} \downarrow_{i} \, \downarrow_{j} } 
\}$, 
reads
\begin{align}
&H_{\mathrm{2Q}}
= \nonumber \\
&
\begin{pmatrix}
\big( \Delta E_{Z} + E_{Z} \big) \hspace*{-3em} & 0 & 0 & 0
\\
0 & \hspace*{-1em} \big( \hspace*{-0.2em} - \hspace*{-0.15em} \Delta E_{Z} + J_{- \, - \, -} + J_{- \, - \, +} \big) \hspace*{-3em} & J_{ij} & 0
\\
0 & J_{ij} & \hspace*{-3.0em} \big( \Delta E_{Z} + J_{+ \, - \, -} + J_{+\,-\,+} \big) \hspace*{-1em} & 0
\\
0 & 0 & 0 & \hspace*{-3em} \big( \hspace*{-0.2em} - \hspace*{-0.15em} \Delta E_{Z} - E_{Z} \big)
\end{pmatrix}.
\label{eq:two_site_one_orbital_exchange_Hamiltonian}
\end{align}
Here, $E_{Z} = E_{Z i}$ is the Zeeman energy of QD $i$, which defines the Zeeman energy difference $\Delta E_{Z} = E_{Z j} - E_{Z i}$, and $\epsilon = \epsilon_{j} - \epsilon_{i}$ is detuning. Furthermore, we have defined
\begin{align}
J_{s_{1} \, s_{2} \, s_{3}} (\epsilon, \Delta E_{Z})
&=
\frac{ t_{ij}^2 }{ s_{1} \Delta E_{Z} + s_{2} \Delta K + s_{3} \epsilon },
\end{align}
where $s_{k} \in \{ -1, +1 \}$, and the exchange strength can be written, as 
\begin{align}
J_{ij}(\epsilon, \Delta E_{Z})
=
\frac{1}{2} \big( 
&
\, J_{+\,+\,+} + J_{+\,+\,-} + J_{-\,+\,+} + J_{-\,+\,-} \big)
\label{eq:perturbative_nearest_neighbour_exchange_strength}
\\
&\approx
\frac{ t^2 }{\Delta K + \epsilon }
+
\frac{ t^2 }{\Delta K - \epsilon },
\nonumber
\\
&=
\frac{ 2 t^2 \Delta K }{ \Delta K^{2} - \epsilon^2 }.
\label{eq:perturbative_nearest_neighbour_exchange_strength_at_small_dEZ}
\end{align}
We note that $J_{ij} \geq 0$. Equation~\eqref{eq:perturbative_nearest_neighbour_exchange_strength_at_small_dEZ} holds when $\Delta E_{Z} \ll \Delta K$. We also define short-hands for the diagonal exchange-like elements
\begin{align}
J_{i}
&=
-J_{-\,-\,-} - J_{-\,-\,+}
\\
J_{j}
&=
-J_{+\,-\,-} - J_{+\,-\,+}.
\end{align}
\subsubsection{Native two-qubit unitary operation}
\label{sec:nearest_neighbour_unitary}
The exact unitary time evolution that $H_{\mathrm{2Q}}(\epsilon)$~\eqref{eq:two_site_one_orbital_exchange_Hamiltonian} generates, which is our native two-qubit interaction, is given by
\begin{align}
&U_{\mathrm{Nat}}(\epsilon)
\nonumber \\
&= \exp(- i H_{\mathrm{2Q}} t / \hbar)
\nonumber \\
&= 
\bigg\{ 
e^{-i (E_{Z} + \Delta E_{Z}) t / \hbar + i J_{ij} t / \hbar}
\ketbra{ \hspace{-0.25em} \uparrow_{i} \uparrow_{j} }{ \uparrow_{i} \uparrow_{j} \hspace{-0.25em} } +
\nonumber \\
&
\big[
\cos(\Omega_{ij} t)
+
\frac{i \Delta_{ij}}{\Omega_{ij}} \sin(\Omega_{ij}t)
\big]
\ketbra{ \hspace{-0.25em} \uparrow_{i} \downarrow_{j} }{ \uparrow_{i} \downarrow_{j} \hspace{-0.25em} } +
\nonumber \\
&
\frac{ i \, J_{ij} }{ \Omega_{ij} } 
\sin ( \Omega_{ij} t )
\big( 
\ketbra{ \hspace{-0.25em} \uparrow_{i} \downarrow_{j} }{ \downarrow_{i} \uparrow_{j} \hspace{-0.25em} }
+
\ketbra{ \hspace{-0.25em} \downarrow_{i} \uparrow_{j} }{ \uparrow_{i} \downarrow_{j} \hspace{-0.25em} } 
\big) +
\nonumber \\
&
\big[
\cos(\Omega_{ij} t)
-
\frac{i \Delta_{ij}}{\Omega_{ij}} \sin(\Omega_{ij}t)
\big]
\ketbra{ \hspace{-0.25em} \downarrow_{i} \uparrow_{j} }{ \downarrow_{i} \uparrow_{j} \hspace{-0.25em} } +
\nonumber \\
&
e^{i (E_{Z} + \Delta E_{Z}) t / \hbar  + i J_{ij} t / \hbar}
\ketbra{ \hspace{-0.25em} \downarrow_{i} \downarrow_{j} }{ \downarrow_{i} \downarrow_{j} \hspace{-0.25em} }
\bigg\} e^{- i J_{ij} t / \hbar} .
\label{eq:U_NNE_single_qubit_parameterisation}
\end{align}
Here, 
\begin{align}
\Delta_{ij}
&:=
\Delta E_{Z} + (J_{i} - J_{j})/2
\\
\hbar \Omega_{ij} (\Delta E_{Z}, J_{ij}) 
&:= 
\sqrt{ \Delta_{ij}^2 + J_{ij}^{2} }.
\label{eq:Omega_DeltaEZ_Jij}
\end{align} 
To analyse the dynamics further, we may parameterise $U_{\mathrm{Nat}}(\epsilon)$, as 
\begin{align}
&U_{\mathrm{Nat}} (\epsilon, \phi, \chi) 
=
e^{i \phi \cos(\chi)} \times 
\nonumber \\
&
\hspace{-1em} \begin{pmatrix}
e^{ -i \varphi_{Z} - i \phi \cos(\chi)} \hspace*{-3.75em} & 0 & 0 & 0
\\
0 
&  
\big[ \cos(\phi) \hspace*{-0.1em} + i \hspace*{-0.1em} \sin(\chi) \sin(\phi) \big]
\hspace*{-1.25em}
& 
-i \cos(\chi) \sin(\phi)
& 
0
\\
0 
& 
-i \cos(\chi) \sin(\phi) 
& 
\hspace*{-1.25em}  
\big[ \cos(\phi) \hspace*{-0.1em} - i \hspace*{-0.1em} \sin(\chi) \sin(\phi) \big]
& 
0
\\
0 & 0 & 0 & \hspace*{-3.75em} e^{ i \varphi_{Z} - i \phi \cos(\chi)}
\end{pmatrix}.
\label{eq:unitary_nearest_neighbour_exchange_phi_chi}
\end{align}
Here, we have defined the parameters
\begin{align}
\chi &:= \mathrm{arctan}(x)
\label{eq:chi_as_arctan_x}
\\
x (\Delta E_{Z}, J_{ij}) 
&:= 
\frac{ \Delta_{ij} }{ J_{ij} }
\label{eq:x_DeltaEZ_Jij_ratio}
\\
\phi(t, \Delta E_{Z}, J_{ij})
&:= 
\frac{ t }{ \Omega_{ij} }.
\label{eq:phi_tau_over_Omega} 
\end{align}
\par
Figure~\ref{fig:x_to_chi_conversion_exact} visualises the conversion between $x$ and $\chi$. Notice that the $x$-axis scale is not linear. 
\documentclass[../../../main/main.tex]{subfiles}
\begin{document}
\begin{figure}
\includegraphics[width=150 bp]{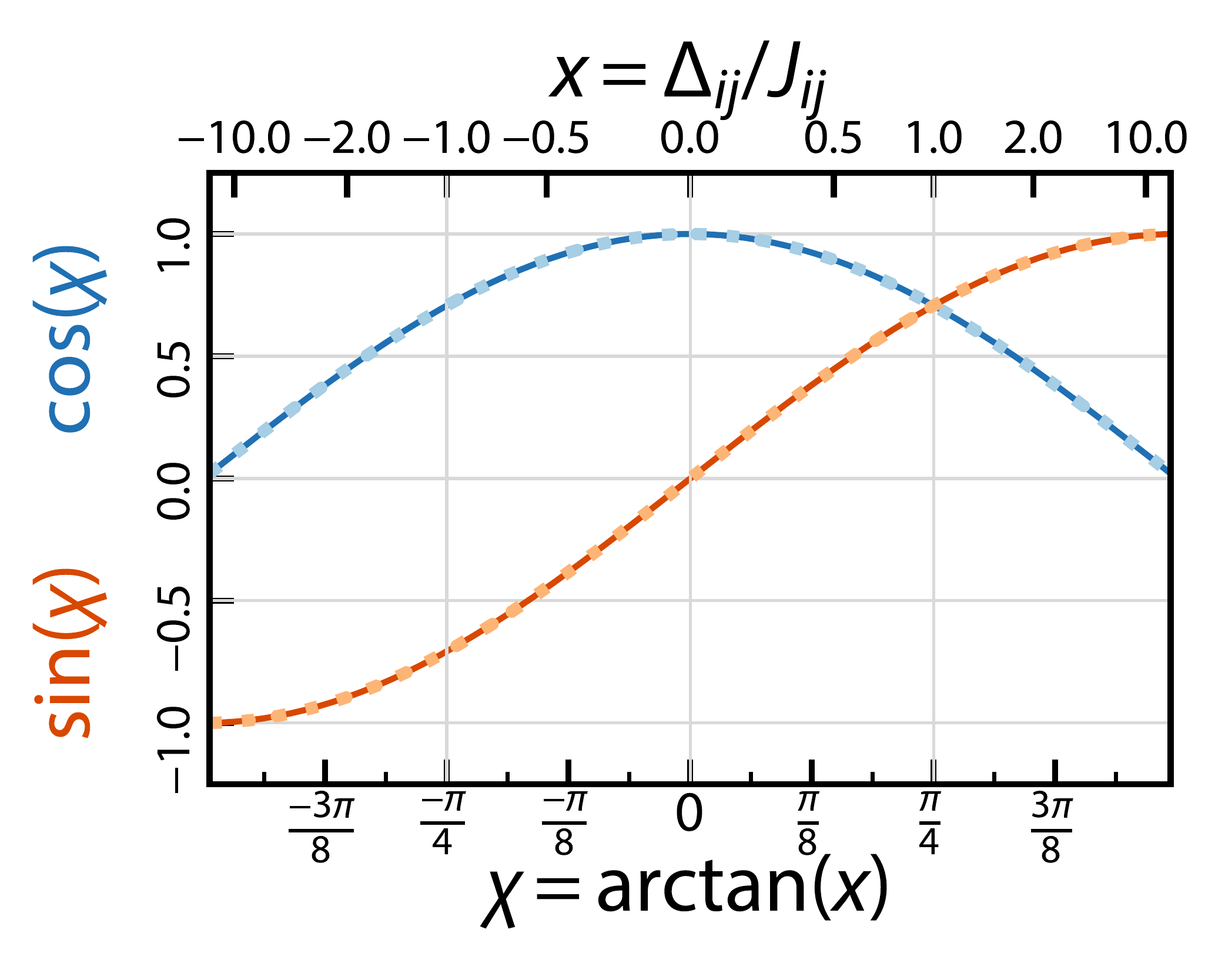}
\caption{
\label{fig:x_to_chi_conversion_exact}
\textbf{Attainable rotation angles.}
Illustration of the required ratio $x = \Delta_{ij} / J_{ij}$ corresponding to the sine (orange trace) and cosine (blue trace) of particular rotation angles $\chi$. 
}
\end{figure}
\end{document}
\subsection{Two-qubit gates}
\label{sec:two_qubit_gates}
We define a few well-known two-qubit gates for reference~\cite{crooks2020gatesSupplement}. From the family of SWAP-rotation gates, 
\begin{align}
\vspace*{-5 bp}
\mathrm{SWAP}
&\,\hat{=}
\begin{pmatrix}
1 & 0 & 0 & 0 \\
0 & 0 & 1 & 0 \\
0 & 1 & 0 & 0 \\
0 & 0 & 0 & 1 
\end{pmatrix} 
\label{eq:SWAP_matrix}
\\
\sqrt{\mathrm{SWAP}}
&\,\hat{=}
\begin{pmatrix}
1 & 0 & 0 & 0 \\
0 & \frac{1+i}{2} & \frac{1-i}{2} & 0 \\
0 & \frac{1-i}{2} & \frac{1+i}{2} & 0 \\
0 & 0 & 0 & 1 
\end{pmatrix}
\label{eq:sqrt_SWAP_matrix}
\\
\mathrm{SWAP}(\theta)
&\hat{=}
\begin{pmatrix}
e^{i \theta} & 0 & 0 & 0 \\
0 & \cos(\theta) & i \sin(\theta) & 0 \\
0 & i \sin(\theta) & \cos(\theta) & 0 \\
0 & 0 & 0 & e^{i \theta} \\
\end{pmatrix} .
\label{eq:SWAP_rotation}
\end{align} 
The so-called Givens rotation gate has a similar form to the SWAP-rotation gate in the $m_{z} = 0$ subspace:  
\begin{align}
\vspace*{-5 bp}
\mathrm{Givens}(\phi)
&\hat{=}
\begin{pmatrix}
1 & 0 & 0 & 0
\\
0 & \sin(\phi) & -\cos(\phi) & 0
\\
0 & \cos(\phi) & \sin(\phi) & 0
\\
0 & 0 & 0 & 1
\end{pmatrix}.
\label{eq:Givens_rotation_gate}
\end{align}
We define the phase gates 
\begin{align}
\vspace*{-5 bp}
\mathrm{CPhase}(\phi)
&\,\hat{=}
\begin{pmatrix}
1 & 0 & 0 & 0 \\
0 & 1 & 0 & 0 \\
0 & 0 & 1 & 0 \\
0 & 0 & 0 & e^{i \phi}
\label{eq:CPhase_gate}
\end{pmatrix}
\\
\mathrm{Ising} (\phi)
&\,\hat{=}
\begin{pmatrix}
1 & 0 & 0 & 0 \\
0 & e^{i \phi} & 0 & 0 \\
0 & 0 & e^{i \phi} & 0 \\
0 & 0 & 0 & 1
\end{pmatrix}.
\end{align}
\subsection{Engineering the native operation}
\label{sec:native_operation_engineering}
Here, we show an example protocol for choosing the exchange strength such, that the resulting unitary time evolution corresponds to the desired two-qubit gate. 
\par
In preconfiguration, the gate time $\tau_{2Q}$, operation (CPhase, Ising, or Givens-like), and rotation angle (either $\alpha$ or $\chi$) are set. In addition, the site-dependent $g$-factors $G_{qi}$ and $G_{qj}$ are known, as well as the detuning $\epsilon$, and external dc magnetic field $B_{0}$. 
\par
As discussed in the main text, the operation determines $\varphi$ such, that for the phase gates we take $\varphi = \pi + 2 \pi n$ for some $n$, and for the Givens-like gate we take $\varphi = \pi/2 + 2 \pi n$. Initially, we assume, that $n = 0$. We then solve for the desired $x$ based on the gate rotation angle, i.e. either $\alpha$ or $\chi$. For the phase gates 
\begin{align}
\alpha
&=
\begin{cases}
\dfrac{\pi + 2 \pi n}{\sqrt{ 1 + x^2 }} + \pi & \mathrm{for\ Ising}
\vspace{4bp} \\
2 \dfrac{\pi + 2 \pi n}{\sqrt{ 1 + x^2 }} & \mathrm{for\ CPhase}.
\end{cases}
\label{eq:alpha_and_x_relationship}
\end{align} 
For the CPhase gate, we take 
\begin{align}
n 
&= 
\mathrm{ceil}(\varphi/\alpha) + 1
\nonumber \\
|x| 
&= 
\dfrac{\sqrt{ (2 \varphi)^2 - \alpha^2 } }{ \sqrt{ \alpha^2 } } .
\nonumber 
\end{align}
For the Ising gate, we take 
\begin{align}
n 
&= 
\mathrm{ceil}( |\varphi|/ | \alpha - \pi |) + 1
\nonumber \\
k 
&= 
\mathrm{floor}(n/2)
\nonumber \\
|x| 
&= 
\dfrac{
\sqrt{\varphi^2 - [(2 k - 1) \pi 
+ \alpha]^2}
}{
(2 k - 1) \pi + \alpha
} .
\nonumber 
\end{align}
For the Givens-like gate, $|x| = \mathrm{tan} (\chi)$ and $n = 0$. The sign of $x$ is determined based on $\Delta E_{Z}$ ($J_{ij} \geq 0$). 
These solutions are not unique.  
We also note that care must be taken if the desired rotation angle $\alpha = 0$ or $\alpha = 2 \pi$ for the CPhase gate, when $\alpha = \pi$ for the Ising gate. 
\par 
The solved $x$ translates to exchange strength $J_{ij}$ (and $\Delta_{ij}$) via $t_{ij}$, and furthermore to $\Omega_{ij}$. Since $\Omega_{ij}$ and $\varphi$ are fixed, in general, the gate time $\tau = \varphi / \Omega_{ij}$ is shorter compared to the target $\tau_{2Q}$. We use the above solutions for $|x|$ at increasing $n$ until $\tau \geq \tau_{2Q}$, to find the $n$ for which $\tau$ is closest to $\tau_{2Q}$. For a high-fidelity operation, we require higher accuracy in $\tau$ than what choosing $n$ can provide. To this end, we may fine-tune $\tau$ using $g$-factor tuning with either of the qubits, to minimize the gate time error
\begin{align}
\delta \tau
&=
\bigg| \tau_{2Q} - \frac{ \varphi }{  \sqrt{[ x^{-1} \Delta E_{Z} (\delta g) ]^2 + \Delta E_{Z}(\delta g) ^2}  \hbar^{-1} } \bigg| .
\end{align}
At the best value of $\delta g$, we re-evaluate the required $t_{ij}$ to hit the desired $J_{ij}$, and hence $x$ and $\Omega_{ij}$. 
\subsection{Semiclassical Rabi model} \label{sec:semiclassical_rabi_model}
The dynamics of a qubit with Larmor frequency $\omega_{0}$, coupling to a transversal magnetic field mode of frequency $\nu$ and amplitude $B_{1}$, leading to coupling strength $g^{*} \mu_{B} B_{1}/\hbar = \omega_{1}$, is described by the semiclassical Rabi model, which we display for convenience. The Hamiltonian reads 
\begin{align}
&H_{1Q} \nonumber 
\\
&= 
\frac{1}{2} \hbar \omega_{0} \sigma_z + \frac{1}{2} \hbar \omega_{1} \big( 
\sigma_{+} e^{-i \nu t + i \varphi}  
+ 
\sigma_{-} e^{i \nu t - i \varphi}  \big) \nonumber
\\
&=
\frac{1}{2} \hbar \omega_{0} \sigma_z 
+ 
\frac{1}{2} \hbar \omega_{1} 
\bigg\{ \sigma_x \cos\big[ \nu t + \varphi \big] 
+ 
\sigma_y \sin\big[ \nu t + \varphi \big] \bigg\},
\label{eq:semiclassical_rabi_hamiltonian}
\end{align}
where $\varphi = \phi + \phi_{\alpha}$.
\par
In a frame rotating by $H_R = e^{- i \omega_{\mathrm{R}} t}$ the semiclassical Rabi Hamiltonian~\eqref{eq:semiclassical_rabi_hamiltonian} retains its form, but the frequencies get replaced, as $\omega_{0} \to \Delta_{0} = \omega_{0} - \omega_{\mathrm{R}}$, $\nu \to \Delta_{\nu} = \nu - \omega_{\mathrm{R}}$. 
\par
The time evolution generated by this Hamiltonian reads
\begin{align}
&U_{1Q}(t, \Delta, \nu)
\nonumber \\
&= \exp( - i H_{1Q} t / \hbar)
\nonumber \\
&=
\frac{e^{i \nu t/2}}{2 \Omega} 
\bigg\{ 
e^{- i \Omega t/2} \big[ \Omega - \Delta \big]
+
e^{  i \Omega t/2} \big[ \Omega + \Delta \big]
\bigg\} 
\ketbra{g}{g}
\nonumber \\
&
- \frac{i \omega_{1}}{\Omega} e^{i \nu t/2 + i \varphi} \sin(\Omega t /2) \ketbra{g}{e}
\nonumber \\
& 
- \frac{i \omega_{1}}{\Omega} e^{-i \nu t/2 - i \varphi} \sin(\Omega t /2) \ketbra{e}{g}
\nonumber \\
&+
\frac{e^{-i \nu t/2}}{2 \Omega} 
\bigg\{ 
e^{- i \Omega t/2} \big[ \Omega + \Delta \big]
+
e^{  i \Omega t/2} \big[ \Omega - \Delta \big]
\bigg\} 
\ketbra{e}{e},
\label{eq:semiclassical_rabi_unitary}
\end{align}
where $\Omega = \sqrt{ \Delta^{2} + \omega_{1}^{2} }$ is the Rabi frequency, and $\Delta = \omega_{0} - \nu$ is the qubit-field detuning. At resonance $\nu = \omega_{0}$, $\Delta = 0$ and $\Omega = \omega_{1}$. 
We may also write the operator in terms of Pauli matrices, as
\begin{align}
&U_{1Q} (t, \Delta, \nu)
\nonumber \\
&=
\big[ \cos(\Omega t/2) \cos(\nu t/2) - \frac{\Delta}{\Omega} \sin (\Omega t/2) \sin(\nu t/2) \big] \textbf{I}
\nonumber \\ 
&
- i \big[ 
\cos(\Omega t/2) \sin(\nu t/2) + \frac{\Delta}{\Omega} \sin (\Omega t/2) \cos(\nu t/2) \big] \sigma_{z} 
\nonumber \\
& +
\frac{ i \omega_{1} }{ \Omega } \sin(\Omega t/2) \big[ \cos(\varphi + \nu t /2) \sigma_{x} + \sin(\varphi + \nu t/2) \sigma_{y} \big]. 
\end{align}
At resonance $\nu = \omega_{0}$, $\Delta = 0$ and $\Omega = \omega_{1}$. Then, 
\begin{align}
&U_{1Q} (t, \Delta = 0, \nu) 
\nonumber \\
&=
\cos(\omega_{1} t/2) \cos(\nu t/2) \textbf{I}
- i 
\cos(\omega_{1} t/2) \sin(\nu t/2) \sigma_{z} 
\nonumber \\
&+
i \sin(\omega_{1} t/2) \big[ \cos(\varphi + \nu t /2) \sigma_{x} + \sin(\varphi + \nu t/2) \sigma_{y} \big]. 
\label{eq:resonant_semiclassical_rabi_unitary_nonresonant_frame}
\end{align}
In the frame $\nu = 0$, the resonant unitary operator (Eq.~\eqref{eq:resonant_semiclassical_rabi_unitary_nonresonant_frame}) reduces to 
\begin{align}
&U_{1Q}(t, \Delta = 0, \nu = 0)
\nonumber \\
&=
\,
\cos(\omega_{1} t/2) \textbf{I}
- i \sin(\omega_{1} t /2)
\big[
e^{i \varphi} \sigma_{-} 
+
e^{- i \varphi} \sigma_{+} 
\big],
\nonumber \\
&=
\cos(\omega_{1} t/2) \textbf{I}
- i \sin(\omega_{1} t /2)
\big[
\cos( \varphi ) \sigma_{x} 
+
\sin( \varphi ) \sigma_{y} 
\big].
\label{eq:resonant_semiclassical_rabi_unitary}
\end{align}
The time evolution Eq.~\eqref{eq:resonant_semiclassical_rabi_unitary} realises $\sqrt{X}$ up to a global phase, as 
\begin{align}
U_{1Q} \big[ t = \pi/(2 \omega_1); \Delta = 0; \nu_{i} = 0 \big] 
&= 
X (\pi/2) \nonumber
\\
&= e^{- i \pi/4} \sqrt{X} .
\end{align}
Here, $X(\theta) = e^{- i \theta \sigma_{x}/2}$. 
\subsection{Multitone driving with frequency binning}
\label{section:frequency_binning}
\documentclass[../../../main/main.tex]{subfiles}
\begin{document}
\begin{figure}
\includegraphics[]{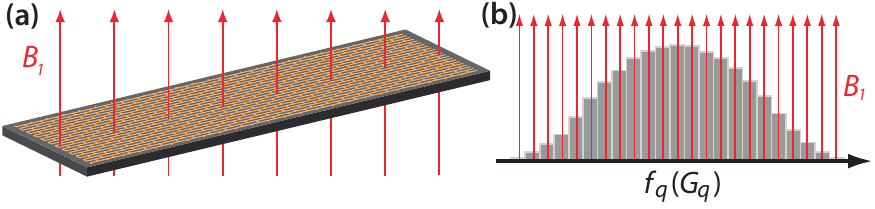}
\caption{
\label{fig:global_sqrtX_gate}
\textbf{Global $\sqrt{X}$ gate.} 
\textbf{(a)} In the nanogrid, we apply transversal single-qubit control (represented as red vertical arrows) chip-globally using e.g. dielectric or superconducting resonators. 
\textbf{(b)} 
Qubits with a Larmor frequency distribution, represented as a gray histogram, can be driven with evenly spaced $B_{1}$ tones, or bins (red vertical arrows). 
Using bin spacing determined by $g$-factor tuneability, $g_{\mathrm{bin}} = 2 \delta g_{\pi}$, all qubits can be tuned to their closest bins. 
}
\end{figure}
\end{document}
In the main text we discuss global transversal control (see Fig.~\ref{fig:global_sqrtX_gate} \textbf{(a)} for illustration). 
The difficulty with single-tone driving is illustrated by the Bloch-sphere polar angle $\mathrm{arccos}( \Delta_{q \nu} / \Omega_{q}(t) )$, where $\Delta_{q \nu} = \omega_{0 q} - \nu$ is the qubit-drive frequency detuning, and $\Omega_{q}(t) = \sqrt{\Delta_{q \nu}^{2} + \omega_{1 q}^{2}(t)}/h$ is the time-dependent Rabi frequency. 
At small detuning $\Delta_{q \nu} \ll \omega_{1 q}(t)$, $\mathrm{arccos}( \Delta_{q \nu} / \Omega_{q} ) \approx \pi/2$ and the effective magnetic field axis is on the transversal plane for all qubits, whereas for $\Delta_{q \nu} \geq \omega_{1 q}(t)$ the effective field axis is significantly qubit-dependent. The small-detuning limit holds when $G_{q}/g_{\mathrm{Si}} \ll B_{1}(t)/B_{0}$. 
At $B_{0} = 1$ T, 
 to reach e.g. $G_{q}/g_{\mathrm{Si}} = 0.05 B_{1}/B_{0}$ with $G_{q} \leq 10^{-2}$, we would require $B_{1} = 100$ mT, 
 which is technologically out of reach. 
 The minimum amplitude required to bring a qubit with $G_{q} = 10^{-2}$ from the Bloch sphere north or south pole onto the transversal plane requires $B_{1} = 5$ mT (coinciding with $\Delta_{q \nu} = \omega_{1q} (t)$). Even this is technologically challenging.  
\par 
While insensitivity to qubit frequencies can be increased with pulse shaping close to the small-detuning limit~\cite{hansen2021pulseSupplement}, far from the small-detuning limit, we propose instead to achieve global control using frequency binning~\cite{fogarty2022siliconSupplement}, which is illustrated in Fig.~\ref{fig:global_sqrtX_gate} \textbf{(b)}. That is, we employ a control pulse of length $\tau_{1Q}$ consisting of $2 N_{1}$ drive tones, at 
\begin{align}
\nu_{i}
&=
\big[ g_{\mathrm{Si}} + i g_{\mathrm{bin}} \big] \mu_{B} B_{0}/ \hbar
\label{eq:frequency_binned_drive_tones}
\end{align}
where $i = 0 \pm 1, 2, 3, ..., N_{1}$ is the bin, and the bin width is determined by the $g$-factor tuneability according to $g_{\mathrm{bin}} = 2 \delta g_{\pi}$, i.e. a bin width of $\nu_{i+1} - \nu_{i} = 2$ MHz. The drive is akin to a finite-component frequency comb. 
Then, using the $g$-factor tuning described above, the frequency of each qubit may be tuned into resonance with the closest bin. 
We expect this bin width to be significantly larger than intrinsic ESR linewidths of $\mathcal{O}(1\ \mathrm{kHz})$~\cite{veldhorst2014addressableSupplement}. For example, choosing $N_{1} = 140$ covers a $g$-factor distribution with $G_{q} \leq \pm 10^{-2}$. 
\par
There is also less power dissipation with frequency binning compared to single-tone driving due to the much smaller required amplitudes. 
At $B_{0} = 1$ T, $B_{1 i} \approx 35.7\ \mu$T yields $\tau_{1Q} = 1\ \upmu$s. 
The dissipated power scales as $P = a |B_{1}|^{2}$ for a constant $a$, such that $a N |B_{1 i}|^{2} = 280 \times |35.7 \times 10^{-6}|^{2} \ll |B_{1}|^{2}$ for $B_{1} = 5$ mT. 
\par
The effect of a single classical drive tone on a qubit is described by the unitary time evolution generated by the semiclassical Rabi model (see Supplementary Sec.~\ref{sec:semiclassical_rabi_model}) $U_{1Q}(t, \Delta_{q \nu}, \nu)$.
While frequency binning allows one to drive all qubits resonantly ($\omega_{0 q} = \nu_i$ for all $q$ for some $i$), viewing the dynamics from a global frame, 
the laboratory frame is not resonant with any of the drive tones. 
In the laboratory frame, qubit dynamics is described by 
\begin{align}
U_{1Q} \big[ t = \pi/(2 \omega_1); \Delta = 0; \nu_{i} \big] 
&=
Z(\varphi_{i}) X( \pi/2)
\label{eq:nonresonant_rabi_unitary}
\end{align}
for some $\varphi_{i}$ which increases with $i$. 
This can be shown, as 
\begin{align}
&Z(\theta) U_{1Q}(\Delta = 0, \nu = 0)
\nonumber \\
&=
\cos(\omega_{1} t/2) \cos(\theta) \textbf{I} - i \cos(\omega_{1} t/2) \sin(\theta) \sigma_{z}   
\nonumber \\
&- i \sin(\omega_{1} t /2) \,
\big[ 
\cos(\varphi + \theta) \sigma_{x} 
+ 
\sin(\varphi + \theta) \sigma_{y}
\big].
\end{align}
Then, $Z(\theta) U_{1Q}(t, \Delta = 0, \nu = 0) = U_{1Q}(t, \Delta = 0, \nu)$ with $\theta = \nu t$. This $Z$-rotation can be absorbed into  $Z(\varphi_{q})$. 
\par
In Fig.~\ref{fig:single_qubit_gate_decomposition}, we visualise gate operations using the decomposition to a rotation direction $(n_{x}, n_{y}, n_{z})$ ($n_{k} \in [-1,1]$), which is a vector on the Bloch sphere, and the rotation angle $\theta \in [ -\pi, \pi ]$. 
The decomposition allows us to study the conformity of $U_{1Q}$ to the $Z(\varphi_{i}) X(\pi/2)$ operation. 
The ideal gate $X(\pi/2)$ has a decomposition with $\theta = \pi/2$, $n_{x} = -1$, $n_{y} = n_{z} = 0$. In comparison, $Z(\varphi_{i}) X(\pi/2)$ has a linearly decreasing $n_{y}$ and $n_{z}$ for linearly increasing $\varphi_{i}$, and thus a small quadratic deviation from $\theta = \pi/2$ and $n_{x} = -1$ with increasing bin number. 
\documentclass[../../../main/main.tex]{subfiles}
\begin{document}
\begin{figure}
\includegraphics[scale=1]{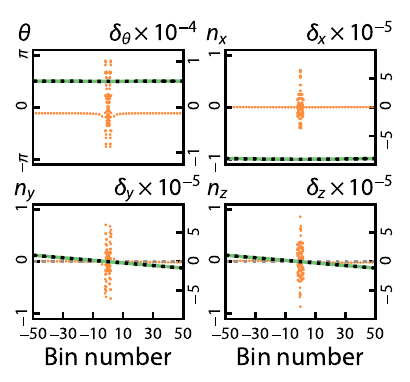}
\caption{
\label{fig:single_qubit_gate_decomposition}
\textbf{Bin-dependence of the column-global single-qubit operation.}
Gate decomposition coefficients $\theta$, $n_{x}$, $n_{y}$, and $n_{z}$ as a function of even bin numbers. Light gray and dark gray dotted lines show the decomposition of ideal $X(\pi/2) = e^{i \pi/4} \sqrt{X}$ and $Z(\varphi_{i}) X(\pi/2)$ for linearly increasing $\varphi_{i}$, respectively. Solid green line shows the decomposition of the
analytical semiclassical Rabi unitary operator with drive tone $\nu_{i}$, and 
numerically integrated time evolution under the semiclassical Rabi Hamiltonian, $U_{1Q}$, with a single drive tone at $\nu_{i}$, corresponding to the resonant bin. Orange datapoints show the decomposition differences $\delta_{\theta}$, $\delta_{x}$, $\delta_{y}$, and $\delta_{z}$ between $U_{1Q}$ with a single drive tone, and five drive tones from $\nu_{i-2}$ to $\nu_{i+2}$ and a smaller $\omega_{1}$. 
The Hamiltonian natively coincides with $X(\pi/2)$ (viewed from the frame rotating with the average $g$-factor Larmor frequency $f_{\mathrm{Si}}$) at resonance $\omega_{0} = \nu$, for qubit Larmor frequency $\omega_{0}$, and drive tone $\nu$, in the reference frame $\nu = 0$ (see the overlap point between the light gray and the blue traces). At reference frames with $\nu \neq 0$, resonantly driven qubit dynamics coincide with $Z(\varphi_{i}) X(\pi/2)$ instead (see the overlap between the dark gray and the green traces). Effects of cross-talk can be compensated with a globally reduced $B_{1}$ amplitude and thus $\omega_{1}$, while retaining a fixed $\tau_{1Q}$ (overlapping solid lines). 
drive tones $\nu_{i}$, and $\nu_{i-1}$, $\nu_{i}$, and $\nu_{i+1}$ for each bin (agreement is better than $1 - 10^{-4}$ for all bins). 
}
\end{figure}
\end{document}
\par
We study the gate fidelity of the analytical semiclassical Rabi unitary $U_{1Q}$ as the conformity to $Z(\varphi_{i}) X(\pi/2)$ under noise. Unless otherwise stated, we use the same parameters as in the simulations described in main text. In particular, we take $B_{1 i} \approx 35.7\ \mu$T (linearly decreasing with increasing bin number), $g_{\mathrm{bin}} = 2 \delta g_{\pi}$. 
The fidelities are essentially bin-independent, since the dominating errors come from the frequency at $f_{\mathrm{Si}}$. 
For a fixed bin $i = 10$, we plot the $Z(\varphi_{i}) X(\pi/2)$ fidelity as a function of gate time error $\sigma_{\tau}$, and noise in the magnetic field component $\sigma_{B_{1}}$, in Fig.~\ref{fig:global_sqrtX_fidelity}. Fidelity behaves similarly to the $Z$-rotation gate fidelity, with noise sources contributing individually. 
\documentclass[../../../main/main.tex]{subfiles}
\begin{document}
\begin{figure}
\includegraphics[]{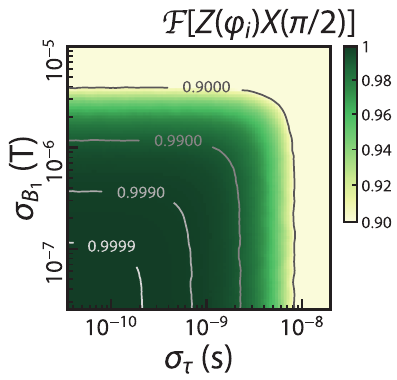}
\caption{
\label{fig:global_sqrtX_fidelity}
\textbf{Global $\sqrt{X}$ gate fidelity.} 
Process fidelity of $Z(\varphi_{i}) X(\pi/2)$ with $Z(\varphi_{i})$ determined by the bin (here, $i = 10$), as a function of variance in gate time $\sigma_{\tau}$, and in the magnetic field component $\sigma_{B_{1}}$. 
}
\end{figure}
\end{document}
\subsection{Variational quantum eigensolvers}
\label{sec:variational_quantum_eigensolvers}
The pipeline can act as the dedicated spin-qubit hardware for the NISQ variational quantum eigensolver (VQE) algorithm of Ref.~\cite{cai2020resourceSupplement}. We thus provide the relationship between the pipeline hardware layout and the VQE application in Ref.~\cite{cai2020resourceSupplement}. 
\par
VQEs try to find the ground state of a problem Hamiltonian $H: \mathcal{H} \to \mathcal{H}$ acting on Hilbert space $\mathcal{H}$ using a circuit that prepares a representation of a Hilbert space state $\ket{ \psi (\pmb{\theta}_{k}) } \in \mathcal{H}$, and measures the expectation value $\bra{ \psi (\pmb{\theta}_{k}) } H \ket{ \psi (\pmb{\theta}_{k}) } $. The state with the smallest expectation value thus provides a best estimate for the ground state energy. 
The parameters $\pmb{\theta}_{k}$ are varied between runs $k = 1,..., N_{\mathrm{r}}$ by varying rotation angles of quantum gates in the circuit. For example, using the so-called Jordan-Wigner mapping, the qubit logical $0$ and $1$ correspond to occupancies of $0$ and $1$ of specific site $s$ and spin $\sigma$, i.e. fermionic Fock states $\ket{ 0_{s \sigma} }$ and $\ket{ 1_{s \sigma} } = c^{\dagger}_{s \sigma} \ket{0}$. In this way, a coherent 50-qubit logical state may represent the coherent superposition of Fock states of $5 \times 5$ fermionic sites. 
\par
A single run of VQE generally consists of four stages: (i) qubit-string state initialisation; (ii) state evolution into a classically solvable Hilbert-space-state; (iii) state evolution into (generally) non-classically-tractable Hilbert-space-state; (iv) as many measurements of observables corresponding to mutually commuting operators, proportional to terms required to measure entire state expectation values of problem Hamiltonian summands (which, summed up, sum to the energy expectation value), as possible. 
The minimum number of runs with fixed parameters for a single estimation of $\bra{ \psi (\pmb{\theta}_{k}) } H \ket{ \psi (\pmb{\theta}_{k}) } $ thus depends on the details of how the problem Hamiltonian summands decompose to commuting (which can be measured during a single run) and non-commuting (requiring different runs) quantum logic gates. 
\par
Stage (iii) contains the variational optimisation, while stage (ii) has fixed-parameter circuit decompositions depending on the classically tractable state of choice. In the so-called Hamiltonian ansatz, stage (iii) consists of a series of blocks of quantum logic gates, each block representing Trotterised time evolution of the Hilbert-space-state, as
\begin{align}
\prod_{n = 1}^{N_{H}}
e^{- i \theta_{n k} h_{n} }
\approx 
\prod_{n = 1}^{N_{H}} (1 - i \theta_{n k} h_{n}),
\end{align}
where the underlying problem Hamiltonian (such as the $5 \times 5$ Fermi-Hubbard Hamiltonian) is the sum
\begin{align}
H
&=
\sum_{n = 1}^{N_{H}} \lambda_{n} h_{n}
\end{align}
with operators $h_{n} : \mathcal{H} \to \mathcal{H}$.
For the Fermi-Hubbard model, the $h_{n}$ are the on-site repulsion operators, and the hopping operators. 
Each $h_{n}$ has fixed gate representations, based on the gate representations of the fermionic operators themselves. The application of the series of these blocks then allows to represent the adiabatic switching on of the on-site repulsion in the simulated model, for example. 
Details of gate decompositions for each stage can be found from Ref.~\cite{cai2020resourceSupplement}. In terms of the hardware implementation, the key results are that all circuits in stages (ii) and (iii) decompose into $Z_{i}(\varphi_{i k})$ and SWAP$_{ij}(\varphi_{ij k})$, laid out in time steps with single- and two-qubit gates alternating in roughly equal numbers, with the two-qubit gates 'weaving' the different qubit states. Stages (i) and (iv) also require stages of $X_{i}(\varphi_{i k})$ and $Y_{i}(\varphi_{i k})$ for the qubit-flips in (i), and for the measurement of observables of mutually commuting operators in (iv). 
\bibliography{references/supplement_references.bib}